\documentclass[11pt]{article}
\usepackage{times}
\usepackage{geometry}
\usepackage[utf8]{inputenc}
\usepackage{enumitem,amssymb}
\usepackage{ragged2e}
\usepackage{graphicx}
\usepackage{comment}
\usepackage{multicol}
\usepackage[usenames]{xcolor} 
\usepackage{wrapfig}

\definecolor{xlinkcolor}{cmyk}{1,1,0,0}
\usepackage{url}
\usepackage[
 colorlinks=true,    
 linkcolor=xlinkcolor,     
 citecolor=xlinkcolor,     
 filecolor=xlinkcolor,  
 urlcolor=xlinkcolor,      
 final=true
]{hyperref}
\usepackage[super,sort&compress]{natbib}
\usepackage{enumitem}
\setenumerate{itemsep=0mm}

\begin{document}
\begin{raggedright} 
\huge
Snowmass2021 - Letter of Interest \hfill \\[+1em]
\textit{Cosmology Intertwined II: The Hubble Constant Tension} \hfill \\[+1em]
\end{raggedright}

\normalsize

\noindent {\large \bf Thematic Areas:}  (check all that apply $\square$/$\blacksquare$)

\noindent $\blacksquare$ (CF1) Dark Matter: Particle Like \\
\noindent $\square$ (CF2) Dark Matter: Wavelike  \\ 
\noindent $\square$ (CF3) Dark Matter: Cosmic Probes  \\
\noindent $\blacksquare$ (CF4) Dark Energy and Cosmic Acceleration: The Modern Universe \\
\noindent $\square$ (CF5) Dark Energy and Cosmic Acceleration: Cosmic Dawn and Before \\
\noindent $\square$ (CF6) Dark Energy and Cosmic Acceleration: Complementarity of Probes and New Facilities \\
\noindent $\blacksquare$ (CF7) Cosmic Probes of Fundamental Physics \\
\noindent $\square$ (Other) {\it [Please specify frontier/topical group]} \\

\noindent {\large \bf Contact Information:}\\
Eleonora Di Valentino (JBCA, University of Manchester, UK) [eleonora.divalentino@manchester.ac.uk]\\

\noindent {\large \bf Authors:}  \\[+1em]
Eleonora Di Valentino (JBCA, University of Manchester, UK)\\
Luis A. Anchordoqui (City University of New York, USA)\\
\"{O}zg\"{u}r Akarsu (Istanbul Technical University, Istanbul, Turkey) \\
Yacine Ali-Haimoud (New York University, USA)\\
Luca Amendola (University of Heidelberg, Germany)\\
Nikki Arendse (DARK, Niels Bohr Institute, Denmark) \\
Marika Asgari (University of Edinburgh, UK)\\
Mario Ballardini (Alma Mater Studiorum Universit\`a di Bologna, Italy)\\
Spyros Basilakos (Academy of Athens and Nat. Observatory of Athens, Greece) \\
Elia Battistelli (Sapienza Universit\`a di Roma and INFN sezione di Roma, Italy)\\
Micol Benetti (Universit\`a degli Studi di Napoli Federico II and INFN sezione di Napoli, Italy)\\
Simon Birrer (Stanford University, USA)\\
Fran\c{c}ois R. Bouchet (Institut d'Astrophysique de Paris, CNRS \& Sorbonne University, France) \\
Marco Bruni (Institute of Cosmology and Gravitation, Portsmouth, UK, and INFN Sezione di Trieste, Italy)\\
Erminia Calabrese (Cardiff University, UK)\\
David Camarena (Federal University of Espirito Santo, Brazil) \\
Salvatore Capozziello (Universit\`a degli Studi di Napoli Federico II, Napoli, Italy) \\
Angela Chen (University of Michigan, Ann Arbor, USA)\\
Jens Chluba (JBCA, University of Manchester, UK)\\
Anton Chudaykin (Institute for Nuclear Research, Russia) \\
Eoin \'O Colg\'ain (Asia Pacific Center for Theoretical Physics, Korea) \\
Francis-Yan Cyr-Racine (University of New Mexico, USA) \\
Paolo de Bernardis (Sapienza Universit\`a di Roma and INFN sezione di Roma, Italy) \\
Javier de Cruz P\'erez (Departament FQA and ICCUB, Universitat de Barcelona, Spain)\\
Jacques Delabrouille (CNRS/IN2P3, Laboratoire APC, France \& CEA/IRFU, France \& USTC, China)\\
Jo Dunkley (Princeton University, USA)\\
Celia Escamilla-Rivera (ICN, Universidad Nacional Aut\'onoma de M\'exico, Mexico) \\
Agn\`es Fert\'e (JPL, Caltech, Pasadena, USA)\\
Fabio Finelli (INAF OAS Bologna and INFN Sezione di Bologna, Italy) \\
Wendy Freedman (University of Chicago, Chicago IL, USA)\\
Noemi Frusciante (Instituto de Astrof\'isica e Ci\^encias do Espa\c{c}o, Lisboa, Portugal)\\
Elena Giusarma (Michigan Technological University, USA) \\
Adri\`a G\'omez-Valent (University of Heidelberg, Germany)\\
Julien Guy (Lawrence Berkeley National Laboratory, USA) \\
Will Handley (University of Cambridge, UK) \\
Ian Harrison (JBCA, University of Manchester, UK) \\
Luke Hart (JBCA, University of Manchester, UK)\\
Alan Heavens (ICIC, Imperial College London, UK)\\
Hendrik Hildebrandt (Ruhr-University Bochum, Germany)\\
Daniel Holz (University of Chicago, Chicago IL, USA)\\
Dragan Huterer (University of Michigan, Ann Arbor, USA)\\
Mikhail M. Ivanov (New York University, USA) \\
Shahab Joudaki (University of Oxford, UK and University of Waterloo, Canada) \\
Marc Kamionkowski (Johns Hopkins University, Baltimore, MD, USA) \\
Tanvi Karwal (University of Pennsylvania, Philadelphia, USA) \\
Lloyd Knox (UC Davis, Davis CA, USA)\\
Suresh Kumar (BITS Pilani, Pilani Campus, India) \\
Luca Lamagna (Sapienza Universit\`a di Roma and INFN sezione di Roma, Italy) \\
Julien Lesgourgues (RWTH Aachen University) \\
Matteo Lucca (Universit\'e Libre de Bruxelles, Belgium)\\
Valerio Marra (Federal University of Espirito Santo, Brazil) \\
Silvia Masi (Sapienza Universit\`a di Roma and INFN sezione di Roma, Italy) \\
Sabino Matarrese (University of Padova and INFN Sezione di Padova, Italy) \\
Arindam Mazumdar (Centre for Theoretical Studies, IIT Kharagpur, India) \\
Alessandro Melchiorri (Sapienza Universit\`a di Roma and INFN sezione di Roma, Italy)\\
Olga Mena (IFIC, CSIC-UV, Spain)\\
Laura Mersini-Houghton (University of North Carolina at Chapel Hill, USA) \\
Vivian Miranda (University of Arizona, USA) \\
Cristian Moreno-Pulido (Departament FQA and ICCUB, Universitat de Barcelona, Spain)\\
David F. Mota (University of Oslo, Norway) \\
Jessica Muir (KIPAC, Stanford University, USA)\\
Ankan Mukherjee (Jamia Millia Islamia Central University, India) \\
Florian Niedermann (CP3-Origins, University of Southern Denmark) \\
Alessio Notari (ICCUB, Universitat de Barcelona, Spain) \\
Rafael C. Nunes (National Institute for Space Research, Brazil)\\
Francesco Pace (JBCA, University of Manchester, UK)\\
Andronikos Paliathanasis (DUT, South Africa and UACh, Chile) \\
Antonella Palmese (Fermi National Accelerator Laboratory, USA) \\
Supriya Pan (Presidency University, Kolkata, India)\\
Daniela Paoletti (INAF OAS Bologna and INFN Sezione di Bologna, Italy)\\
Valeria Pettorino (AIM, CEA, CNRS, Universit\'e Paris-Saclay, Universit\'e de Paris, France) \\
Francesco Piacentini (Sapienza Universit\`a di Roma and INFN sezione di Roma, Italy)\\
Vivian Poulin (LUPM, CNRS \& University of Montpellier, France) \\
Marco Raveri (University of Pennsylvania, Philadelphia, USA) \\
Adam G. Riess (Johns Hopkins University, Baltimore, USA) \\
Vincenzo Salzano (University of Szczecin, Poland)\\
Emmanuel N. Saridakis (National Observatory of Athens, Greece)\\
Anjan A. Sen (Jamia Millia Islamia Central University New Delhi, India) \\
Arman Shafieloo (Korea Astronomy and Space Science Institute (KASI), Korea)\\
Anowar J. Shajib (University of California, Los Angeles, USA) \\
Joseph Silk (IAP Sorbonne University \& CNRS, France, and Johns Hopkins University, USA)\\
Alessandra Silvestri (Leiden University, NL)\\
Martin S. Sloth (CP3-Origins, University of Southern Denmark) \\
Tristan L. Smith (Swarthmore College, Swarthmore, USA)\\ 
Joan Sol\`a Peracaula (Departament FQA and ICCUB, Universitat de Barcelona, Spain)\\
Carsten van de Bruck (University of Sheffield, UK) \\
Licia Verde (ICREA, Universidad de Barcelona, Spain)\\
Luca Visinelli (GRAPPA, University of Amsterdam, NL) \\
Benjamin D. Wandelt (IAP Sorbonne University \& CNRS, France, and CCA, USA) \\
Deng Wang (National Astronomical Observatories, CAS, China) \\
Jian-Min Wang (Key Laboratory for Particle Astrophysics, IHEP of the CAS, Beijing, China) \\
Anil K. Yadav (United College of Engg. \& Research, GN, India)\\
Weiqiang Yang (Liaoning Normal University, Dalian, China) \\

\noindent {\large \bf Abstract:} 
The current cosmological probes have provided a fantastic confirmation of the standard $\Lambda$ Cold Dark Matter cosmological model, that has been constrained with unprecedented accuracy. However, with the increase of the experimental sensitivity a few statistically significant tensions between different independent cosmological datasets emerged. While these tensions can be in portion the result of systematic errors, the persistence after several years of accurate analysis strongly hints at cracks in the standard cosmological scenario and the need for new physics. In this Letter of Interest we will focus on the $4.4\sigma$ tension between the Planck estimate of the Hubble constant $H_0$ and the SH0ES collaboration measurements. After showing the $H_0$ evaluations made from different teams using different methods and geometric calibrations, we will list a few interesting new physics models that could solve this tension and discuss how the next decade experiments will be crucial.

\clearpage

\noindent {\bf State-of-the-art --} The 2018 legacy release from the Planck satellite \cite{Akrami:2018vks} of the Cosmic Microwave Background (CMB) anisotropies, has provided a fantastic confirmation of the standard $\Lambda$ Cold Dark Matter ($\Lambda$CDM) cosmological model. However, the improvement in estimating the uncertainties has led to statistically-significant tensions in the measurement of various quantities between Planck and independent cosmological probes. While some proportion of these discrepancies may have a systematic origin, their magnitude and persistence across probes strongly hint at cracks in the standard cosmological scenario and the need for new physics. The most statistically significant tension is in the estimation of the {\it Hubble constant} $H_0$ between the CMB, assuming a $\Lambda$CDM model, and the direct local distance ladder measurements. In particular, the Planck collaboration~\cite{Aghanim:2018eyx} finds  $H_0=\left(67.27\pm0.60\right)$ km/s/Mpc\footnote{All the bounds are reported at 68\% confidence level in the text.}. This constraint is in tension at about $4.4\sigma$ with the 2019 SH0ES collaboration (R19~\cite{Riess:2019cxk}) constraint, $H_0=(74.03 \pm 1.42)$ km/s/Mpc, based on the analysis of the Hubble Space Telescope observations using 70 long-period Cepheids in the Large Magellanic Cloud.

\begin{wrapfigure}{R}{0.36\textwidth}
\centering
\includegraphics[width=0.35\textwidth]{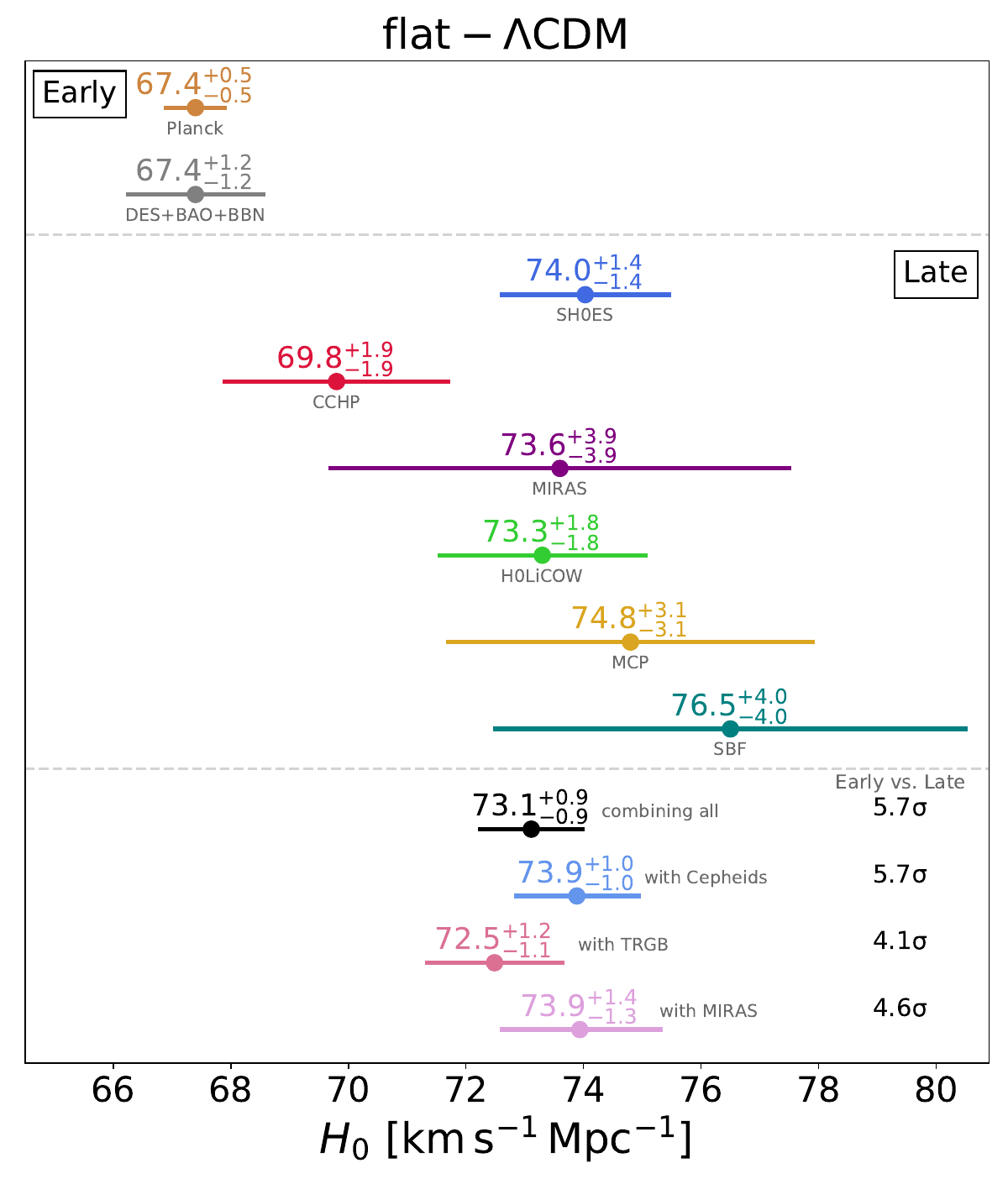}
\caption{68\% CL constraint on $H_0$ from different cosmological probes (from Ref.~\cite{Verde:2019ivm}).}
\label{whisker}
\end{wrapfigure}

As shown in Fig.~\ref{whisker}, preferring smaller values, we have the early universe estimates of $H_0$, as obtained by Planck or by ACT+WMAP~\cite{Aiola:2020azj} ($H_0=(67.6 \pm 1.1)$ km/s/Mpc), and their combination with Baryon Acoustic Oscillation (BAO) data~\cite{Beutler:2011hx,Ross:2014qpa,Alam:2016hwk}, the Y1 measurements of the Dark Energy Survey~\cite{Troxel:2017xyo, Abbott:2017wau, Krause:2017ekm}, supernovae from the Pantheon catalog~\cite{Scolnic:2017caz}, and a prior on the baryon density derived from measurements of primordial deuterium~\cite{Cooke:2017cwo} assuming standard Big Bang Nucleosynthesis (BBN). A reanalysis of the BOSS full-shape data~\cite{Ivanov:2019pdj,DAmico:2019fhj}, as well as BAO+BBN~\cite{Alam:2020sor} from BOSS and eBOSS provides $H_0=(67.35 \pm 0.97)$, while SPTpol~\cite{Henning:2017nuy} finds $H_0=(71.3 \pm 2.1)$ km/s/Mpc. In contrast, standard distance ladder and time delay distances agree on a low-$z$ high-$H_0$ value, as the SH0ES estimate~\cite{Reid:2019tiq} $H_0=(73.5 \pm 1.4)$ km/s/Mpc, and the H0LiCOW~\cite{Wong:2019kwg} inferred value $H_0=(73.3^{+1.7}_{-1.8})$ km/s/Mpc, based on strong gravitational lensing effects on quasar systems. However, the strong lensing TDCOSMO+SLACS~\cite{Birrer:2020tax} sample prefers $H_0=67.4 ^{+4.1}_{-3.2}$ km/s/Mpc.
Then, we have the reanalysis of the Cepheid data by using Bayesian hyper-parameters~\cite{Cardona:2016ems}, the local determination of $H_0$~\cite{Camarena:2019moy} considering the cosmographic expansion of the luminosity distance, the independent determination of $H_0$ based on the Tip of the Red Giant Branch~\cite{Freedman:2019jwv,Yuan:2019npk,Freedman:2020dne}, and that obtained by using the Surface Brightness Fluctuations method~\cite{Verde:2019ivm,Khetan:2020hmh}, or the Cosmic Chronometers~\cite{Yu:2017iju,Gomez-Valent:2019lny,Haridasu:2018gqm,Dutta:2019pio}. Finally, a larger value for $H_0$ is preferred by MIRAS~\cite{Huang:2019yhh} (variable red giant stars), by STRIDES~\cite{Shajib:2019toy}, using the Infrared~\cite{Kourkchi:2020iyz} or Baryonic Tully–Fisher relation~\cite{Schombert:2020pxm}, or by Standardized Type II supernovae~\cite{deJaeger:2020zpb}. There is no single type of systematic measurement error in Cepheids which could solve the $H_0$ crisis, as speculated in~\cite{Efstathiou:2020wxn} (e.g., it would not work for Cepheids calibrated in NGC 4258), and in any case it could not explain the final result from the Maser Cosmology Project~\cite{Pesce:2020xfe}, completely independent from these considerations, that finds $H_0=(73.9 \pm 3.0)$ km/s/Mpc. If the late universe estimates are averaged in different combinations, these $H_0$ values disagree between 4.5$\sigma$ and 6.3$\sigma$ with those from Planck~\cite{Riess:2020sih}.

\noindent {\bf Possible solutions --} Models addressing the $H_0$ tension are extremely difficult to concoct. 
The simplest possibility is a sample-variance effect, due to an underdense local universe. However, this is a factor of $\sim 20$ too small to explain the $H_0$ tension, and thus decisively ruled out~\cite{Wu:2017fpr,Kenworthy:2019qwq}. This leaves a host of many proposed partial explanations~\cite{Tram:2016rcw,DiValentino:2016ziq,DiValentino:2016hlg,Bernal:2016gxb,DiValentino:2017oaw,DiValentino:2017rcr,Binder:2017lkj,Khosravi:2017hfi,DiValentino:2017zyq,Renk:2017rzu,DiValentino:2017gzb,Kumar:2016zpg,DiValentino:2017iww,Kumar:2017dnp,Gomez-Valent:2020mqn,Lucca:2020zjb,vandeBruck:2017idm,Yang:2018euj,Yang:2018uae,Yang:2019uzo,Martinelli:2019dau,DiValentino:2019ffd,DiValentino:2019jae,Benevento:2020fev,Belgacem:2017cqo,Fernandez-Arenas:2017isq,Sola:2017znb,Nunes:2018xbm,Colgain:2018wgk,DEramo:2018vss,Guo:2018ans,Lin:2018nxe,Yang:2018qmz,Vagnozzi:2019ezj,DiValentino:2019dzu,DiValentino:2020naf,Keeley:2019esp,Joudaki:2016kym,Yang:2018prh,Poulin:2018cxd,Karwal:2016vyq,Banihashemi:2018oxo,Banihashemi:2018has,Mortsell:2018mfj,Zhang:2018air,Kreisch:2019yzn,Kumar:2019wfs,Agrawal:2019lmo,Yang:2019jwn,Yang:2019qza,DiValentino:2019exe,Desmond:2019ygn,Yang:2019nhz,Pan:2019gop,Visinelli:2019qqu,Smith:2019ihp,Lucca:2020fgp,Lin:2019qug,Martinelli:2019krf,Cai:2019bdh,Schoneberg:2019wmt,Shafieloo:2016bpk,Li:2019san,Cuceu:2019for,Colgain:2019joh,Pan:2019jqh,Berghaus:2019cls,Knox:2019rjx,Pandey:2019plg,Adhikari:2019fvb,Lancaster:2017ksf,Niedermann:2019olb,Yadav:2019jio,Kasai:2019yqn,Amirhashchi:2020qep,Perez:2020cwa,Pan:2020bur,DAgostino:2020dhv,Liao:2020zko,Yang:2020zuk,Pan:2020zza,Wu:2020nxz,Blinov:2020hmc,Wang:2020zfv,Chudaykin:2020acu,Alestas:2020mvb,Clark:2020miy,Keeley:2020rmo,Niedermann:2020dwg,Archidiacono:2020yey,DiValentino:2020kha,Capozziello:2020nyq,Anchordoqui:2019yzc,Ivanov:2020mfr,Gonzalez:2020fdy,Hryczuk:2020jhi,Baxter:2020qlr,Anchordoqui:2011nh,Jacques:2013xr,Weinberg:2013kea,Anchordoqui:2012qu,Carneiro:2018xwq,Paul:2018njm,DiValentino:2015sam,Green:2019glg,Barenboim:2016lxv,Ferreira:2018vjj,Gelmini:2019deq,DiValentino:2015wba,Poulin:2018dzj,Baumann:2016wac,Anchordoqui:2020znj,Sakstein:2019fmf,Das:2020wfe,Akarsu:2019hmw,Ye:2020btb,Hart:2017ndk, Chiang:2018xpn, Hart:2019dxi, Jedamzik:2020krr, Sekiguchi:2020teg,Bose:2020cjb,Agrawal:2019dlm,Anchordoqui:2019amx,Anchordoqui:2020sqo,Ade:2015rim,Raveri:2019mxg,Yan:2019gbw,Frusciante:2019puu,Braglia:2020iik,Ballardini:2020iws,Rossi:2019lgt,Pan:2019hac,Li:2019yem,Rezaei:2020mrj,Liu:2020vgn,Li:2020ybr,Yang:2020tax,DiBari:2013dna,Choi:2019jck,Choi:2020tqp,Berezhiani:2015yta,Anchordoqui:2015lqa,Vattis:2019efj,Desai:2019pvs,Alcaniz:2019kah,Chudaykin:2016yfk,Chudaykin:2017ptd,Hill:2020osr,Ivanov:2020ril,Rezaei:2020lfy,Wang:2020dsc,Nunes:2020uex,Leonhardt:2020qam,Birrer:2020jyr,Ballesteros:2020sik,Blinov:2019gcj,Hernandez-Almada:2020uyr,Philcox:2020xbv,Feeney:2017sgx,Feeney:2018mkj,Mortlock:2018azx,Banerjee:2020xcn,Adler:2019fnp,Gu:2020ozv,Akarsu:2019pwn}, but none of them offer a fully satisfactory solution when all other data and parameters are taken into account
~\cite{DiValentino:2020vhf,DiValentino:2020vvd,DiValentino:2020srs}. The models can have a \textcolor{blue}{dark energy (DE)} explanation or \textcolor{red}{not}:

\begin{itemize}[noitemsep,topsep=0pt]

\item [\textcolor{blue}{\textbullet}] A DE component with an equation of state $w \neq -1$, i.e. allowing for deviation from the cosmological constant $\Lambda$, both constant or dynamical with redshift~\cite{Aghanim:2018eyx,Yang:2018qmz,Yang:2018prh,DiValentino:2019dzu,Vagnozzi:2019ezj,DiValentino:2020naf,Keeley:2019esp,Joudaki:2016kym}. These models usually solve the $H_0$ tension within two standard deviations at the price of a phantom-like DE, i.e. $w<-1$,  because of the geometrical degeneracy present with the DE equation of state $w$.

\item [\textcolor{blue}{\textbullet}] Early dark energy (EDE) which behaves like $\Lambda$ at $z \geq 3000$ and decays away as radiation or faster at later times~\cite{Pettorino:2013ia,Poulin:2018cxd,Karwal:2016vyq}. Related models include: {\it (i)}~coupling of the EDE scalar to neutrinos~\cite{Sakstein:2019fmf}; {\it (ii)}~a first-order phase transition in a dark sector before recombination which leads to a short phase of EDE~\cite{Niedermann:2019olb};  {\it (iii)}~an EDE model with an Anti-de Sitter phase around recombination~\cite{Akarsu:2019hmw,Ye:2020btb}; {\it (iv)}~an evolving scalar field asymptotically oscillating or with a non-canonical kinetic term~\cite{Agrawal:2019lmo,Lin:2019qug}, {\it (v)}~an axion-like particle sourcing dark radiation~\cite{Berghaus:2019cls}, {\it (vi)}~a scalar field with a potential inspired by ultra-light axions~\cite{Smith:2019ihp,Lucca:2020fgp}.

\item [\textcolor{blue}{\textbullet}] Interacting dark energy (IDE) models, where dark matter (DM) and DE share interactions other than gravitational~\cite{Pettorino:2013oxa,Kumar:2016zpg,DiValentino:2017iww,Kumar:2017dnp,vandeBruck:2017idm,Yang:2018euj,Yang:2018uae,Yang:2019uzo,Martinelli:2019dau, DiValentino:2019ffd,DiValentino:2019jae,Benevento:2020fev,Gomez-Valent:2020mqn,Lucca:2020zjb,Yang:2020uga,Yang:2019uog,Yang:2018ubt}. The IDE model solves the tension with R19 within one standard deviation, leading to a preference for a non-zero DE-DM coupling at more than $5$ standard deviations~\cite{DiValentino:2019ffd,DiValentino:2019jae}, fixing the DE equation of state to a cosmological constant.
However, this category can be further extended into two classes~\cite{DiValentino:2019jae}: {\it (i)}~models with $w < - 1$ in which energy flows from DE to DM, {\it (ii)}~models with $w> -1$ in which energy flows from DM to DE. Related models can be realized in string theory~\cite{Agrawal:2019dlm,Anchordoqui:2019amx,Anchordoqui:2020sqo}.

\item [\textcolor{blue}{\textbullet}] Phenomenologically Emergent Dark Energy~\cite{Li:2019yem,Pan:2019hac,Rezaei:2020mrj,Liu:2020vgn,Li:2020ybr,Yang:2020tax}, where the $H_0$ tension with R19 is alleviated within one standard deviation without additional degrees of freedom with respect to $\Lambda$CDM.

\item  [\textcolor{red}{\textbullet}] Extra relativistic degrees of freedom at recombination, parametrized by the number of equivalent light neutrino species $N_{\rm eff}$~\cite{Steigman:1977kc}. For three active massless neutrino families, $N_{\rm eff}^{\rm SM} \simeq 3.046$~\cite{Mangano:2005cc,deSalas:2016ztq,Akita:2020szl}.  For the well-known degeneracy, we can increase $H_0$ at the price of additional radiation at recombination. Sterile neutrinos, Goldstone bosons, axions, and neutrino asymmetry are typical examples to enhance the value of $N_{\rm eff}$~\cite{Anchordoqui:2011nh,Jacques:2013xr,Weinberg:2013kea,Anchordoqui:2012qu,Carneiro:2018xwq,Paul:2018njm,DiValentino:2015sam,Green:2019glg,Ferreira:2018vjj,Gelmini:2019deq,DiValentino:2015wba,Poulin:2018dzj,Baumann:2016wac,Barenboim:2016lxv,Zeng:2018pcv,Allahverdi:2014ppa}. Future surveys will detect deviations from $N_{\rm eff}^{\rm SM}$ within $\Delta N_{\rm eff} \lesssim 0.06$ at 95\% CL, allowing to probe a vast range of light relic models~\cite{Abazajian:2016yjj, Abazajian:2019eic}.

\item [\textcolor{red}{\textbullet}] Modified recombination and reionization histories through heating processes, variation of fundamental constants, or a non-standard CMB temperature-redshift relation \citep{Hart:2017ndk, Chiang:2018xpn, Hart:2019dxi, Jedamzik:2020krr, Sekiguchi:2020teg,Bose:2020cjb}.

\item [\textcolor{red}{\textbullet}] Modified Gravity models~\cite{Ade:2015rim} in which gravity changes with redshift, such that the $H_0$ estimate from CMB can have larger values~\cite{Raveri:2019mxg,Yan:2019gbw,Frusciante:2019puu,Sola:2019jek,Sola:2020lba,Braglia:2020iik,Ballardini:2020iws,Rossi:2019lgt,Umilta:2015cta,Ballardini:2016cvy}.

\item [\textcolor{red}{\textbullet}] Decaying dark matter~\cite{DiBari:2013dna,Choi:2019jck,Choi:2020tqp,Berezhiani:2015yta,Anchordoqui:2015lqa,Vattis:2019efj,Desai:2019pvs,Alcaniz:2019kah,Chudaykin:2016yfk,Chudaykin:2017ptd} or interacting neutrinos~\cite{Kreisch:2019yzn,Blinov:2019gcj,DiValentino:2017oaw}.

\end{itemize}

\noindent Theoretical efforts to find a dynamic model describing the data have been placed side by side to kinematic models, as the cosmography, where the current expansion is a function of the cosmic time~\cite{Sahni:2014ooa, Capozziello:2019cav, Benetti:2019gmo}.

\noindent {\bf Standard Sirens --} In the next decade an important role will be played by standard sirens (GWSS)~\cite{Schutz:1986gp,Holz:2005df,Chen:2017rfc,DiValentino:2018jbh,Palmese:2019ehe}, the gravitational-wave (GW) analog of astronomical standard candles.
In fact, the observations of the merger of the binary neutron-star system GW170817~\cite{Abbott:2017xzu} provided $H_0=70_{-\,8}^{+12}$ km/s/Mpc. While this constraint is significantly relaxed, it does not require any form of cosmic ‘distance ladder’ and it is model-independent. It can be important in an extended parameter space~\cite{DiValentino:2017clw} in which CMB data are unable to strongly constrain $H_0$.
At least 25 additional observations of GWSS~\cite{Nissanke:2013fka} are needed to discriminate between Planck and R19. An uncertainty of $1-2\%$ in $H_0$ is expected in the early(mid)-2020s~\cite{Chen:2017rfc}, from the analysis of GW events with electromagnetic counterparts. Finally, complementary dark GWSS, as the GW190814 in~\cite{Palmese:2020aof}, are expected to provide a $1-4\%$ constraint on $H_0$ using the second generation of the detector networks~\cite{Yu:2020vyy,Borhanian:2020vyr}. 

\noindent {\bf Looking into the future --} Solving the $H_0$ tension is very much an ongoing enterprise. The resolution of this conundrum will likely require a coordinated effort from the side of theory and interpretation (providing crucial tests of the exotic cosmologies), and data analysis and observation (expected to improve methods and disentangle systematics). This agenda will flourish in the next decade with future CMB experiments, as the Simon Observatory or CMB-S4, that combined with gigantic cosmic surveys, as Euclid and LSST, are expected to reach an uncertainty of $\sim 0.15\%$ in the $H_0$ estimate. In summary, the next decade will test the $\Lambda$CDM model and build the next-generation experiments that will usher in a new era of cosmology.

\clearpage

\bibliographystyle{utphys}
\bibliography{H0}

\providecommand{\href}[2]{#2}\begingroup\raggedright\begin{thebibliography}{100}

\bibitem{Akrami:2018vks}
{\bfseries Planck} Collaboration, Y.~Akrami {\em et~al.}, ``{Planck 2018
  results. I. Overview and the cosmological legacy of Planck},''
\href{http://arxiv.org/abs/1807.06205}{{\ttfamily arXiv:1807.06205
  [astro-ph.CO]}}.

\bibitem{Aghanim:2018eyx}
{\bfseries Planck} Collaboration, N.~Aghanim {\em et~al.}, ``{Planck 2018
  results. VI. Cosmological parameters},''
\href{http://arxiv.org/abs/1807.06209}{{\ttfamily arXiv:1807.06209
  [astro-ph.CO]}}.

\bibitem{Riess:2019cxk}
A.~G. Riess, S.~Casertano, W.~Yuan, L.~M. Macri, and D.~Scolnic, ``{Large
  Magellanic Cloud Cepheid Standards Provide a 1\% Foundation for the
  Determination of the Hubble Constant and Stronger Evidence for Physics beyond
  $\Lambda$CDM},'' \href{http://dx.doi.org/10.3847/1538-4357/ab1422}{{\em
  Astrophys. J.} {\bfseries 876} no.~1, (2019) 85},
\href{http://arxiv.org/abs/1903.07603}{{\ttfamily arXiv:1903.07603
  [astro-ph.CO]}}.

\bibitem{Verde:2019ivm}
L.~Verde, T.~Treu, and A.~Riess, ``{Tensions between the Early and the Late
  Universe},'' \href{http://arxiv.org/abs/1907.10625}{{\ttfamily
  arXiv:1907.10625 [astro-ph.CO]}}.

\bibitem{Aiola:2020azj}
{\bfseries ACT} Collaboration, S.~Aiola {\em et~al.}, ``{The Atacama Cosmology
  Telescope: DR4 Maps and Cosmological Parameters},''
  \href{http://arxiv.org/abs/2007.07288}{{\ttfamily arXiv:2007.07288
  [astro-ph.CO]}}.

\bibitem{Beutler:2011hx}
F.~Beutler, C.~Blake, M.~Colless, D.~H. Jones, L.~Staveley-Smith, L.~Campbell,
  Q.~Parker, W.~Saunders, and F.~Watson, ``{The 6dF Galaxy Survey: Baryon
  Acoustic Oscillations and the Local Hubble Constant},''
  \href{http://dx.doi.org/10.1111/j.1365-2966.2011.19250.x}{{\em Mon. Not. Roy.
  Astron. Soc.} {\bfseries 416} (2011) 3017--3032},
\href{http://arxiv.org/abs/1106.3366}{{\ttfamily arXiv:1106.3366
  [astro-ph.CO]}}.

\bibitem{Ross:2014qpa}
A.~J. Ross, L.~Samushia, C.~Howlett, W.~J. Percival, A.~Burden, and M.~Manera,
  ``{The clustering of the SDSS DR7 main Galaxy sample – I. A 4 per cent
  distance measure at $z = 0.15$},''
  \href{http://dx.doi.org/10.1093/mnras/stv154}{{\em Mon. Not. Roy. Astron.
  Soc.} {\bfseries 449} no.~1, (2015) 835--847},
\href{http://arxiv.org/abs/1409.3242}{{\ttfamily arXiv:1409.3242
  [astro-ph.CO]}}.

\bibitem{Alam:2016hwk}
{\bfseries BOSS} Collaboration, S.~Alam {\em et~al.}, ``{The clustering of
  galaxies in the completed SDSS-III Baryon Oscillation Spectroscopic Survey:
  cosmological analysis of the DR12 galaxy sample},''
  \href{http://dx.doi.org/10.1093/mnras/stx721}{{\em Mon. Not. Roy. Astron.
  Soc.} {\bfseries 470} no.~3, (2017) 2617--2652},
\href{http://arxiv.org/abs/1607.03155}{{\ttfamily arXiv:1607.03155
  [astro-ph.CO]}}.

\bibitem{Troxel:2017xyo}
{\bfseries DES} Collaboration, M.~A. Troxel {\em et~al.}, ``{Dark Energy Survey
  Year 1 results: Cosmological constraints from cosmic shear},''
  \href{http://dx.doi.org/10.1103/PhysRevD.98.043528}{{\em Phys. Rev.}
  {\bfseries D98} no.~4, (2018) 043528},
\href{http://arxiv.org/abs/1708.01538}{{\ttfamily arXiv:1708.01538
  [astro-ph.CO]}}.

\bibitem{Abbott:2017wau}
{\bfseries DES} Collaboration, T.~M.~C. Abbott {\em et~al.}, ``{Dark Energy
  Survey year 1 results: Cosmological constraints from galaxy clustering and
  weak lensing},'' \href{http://dx.doi.org/10.1103/PhysRevD.98.043526}{{\em
  Phys. Rev.} {\bfseries D98} no.~4, (2018) 043526},
\href{http://arxiv.org/abs/1708.01530}{{\ttfamily arXiv:1708.01530
  [astro-ph.CO]}}.

\bibitem{Krause:2017ekm}
{\bfseries DES} Collaboration, E.~Krause {\em et~al.}, ``{Dark Energy Survey
  Year 1 Results: Multi-Probe Methodology and Simulated Likelihood Analyses},''
  \href{http://arxiv.org/abs/1706.09359}{{\ttfamily arXiv:1706.09359
  [astro-ph.CO]}}.

\bibitem{Scolnic:2017caz}
D.~M. Scolnic {\em et~al.}, ``{The Complete Light-curve Sample of
  Spectroscopically Confirmed SNe Ia from Pan-STARRS1 and Cosmological
  Constraints from the Combined Pantheon Sample},''
  \href{http://dx.doi.org/10.3847/1538-4357/aab9bb}{{\em Astrophys. J.}
  {\bfseries 859} no.~2, (2018) 101},
\href{http://arxiv.org/abs/1710.00845}{{\ttfamily arXiv:1710.00845
  [astro-ph.CO]}}.

\bibitem{Cooke:2017cwo}
R.~J. Cooke, M.~Pettini, and C.~C. Steidel, ``{One Percent Determination of the
  Primordial Deuterium Abundance},''
  \href{http://dx.doi.org/10.3847/1538-4357/aaab53}{{\em Astrophys. J.}
  {\bfseries 855} no.~2, (2018) 102},
\href{http://arxiv.org/abs/1710.11129}{{\ttfamily arXiv:1710.11129
  [astro-ph.CO]}}.

\bibitem{Ivanov:2019pdj}
M.~M. Ivanov, M.~Simonovi\'c, and M.~Zaldarriaga, ``{Cosmological Parameters
  from the BOSS Galaxy Power Spectrum},''
  \href{http://dx.doi.org/10.1088/1475-7516/2020/05/042}{{\em JCAP} {\bfseries
  05} (2020) 042}, \href{http://arxiv.org/abs/1909.05277}{{\ttfamily
  arXiv:1909.05277 [astro-ph.CO]}}.

\bibitem{DAmico:2019fhj}
G.~D'Amico, J.~Gleyzes, N.~Kokron, D.~Markovic, L.~Senatore, P.~Zhang,
  F.~Beutler, and H.~Gil-Marín, ``{The Cosmological Analysis of the SDSS/BOSS
  data from the Effective Field Theory of Large-Scale Structure},''
  \href{http://dx.doi.org/10.1088/1475-7516/2020/05/005}{{\em JCAP} {\bfseries
  05} (2020) 005}, \href{http://arxiv.org/abs/1909.05271}{{\ttfamily
  arXiv:1909.05271 [astro-ph.CO]}}.

\bibitem{Alam:2020sor}
{\bfseries eBOSS} Collaboration, S.~Alam {\em et~al.}, ``{The Completed SDSS-IV
  extended Baryon Oscillation Spectroscopic Survey: Cosmological Implications
  from two Decades of Spectroscopic Surveys at the Apache Point observatory},''
  \href{http://arxiv.org/abs/2007.08991}{{\ttfamily arXiv:2007.08991
  [astro-ph.CO]}}.

\bibitem{Henning:2017nuy}
{\bfseries SPT} Collaboration, J.~Henning {\em et~al.}, ``{Measurements of the
  Temperature and E-Mode Polarization of the CMB from 500 Square Degrees of
  SPTpol Data},'' \href{http://dx.doi.org/10.3847/1538-4357/aa9ff4}{{\em
  Astrophys. J.} {\bfseries 852} no.~2, (2018) 97},
  \href{http://arxiv.org/abs/1707.09353}{{\ttfamily arXiv:1707.09353
  [astro-ph.CO]}}.

\bibitem{Reid:2019tiq}
M.~Reid, D.~Pesce, and A.~Riess, ``{An Improved Distance to NGC 4258 and its
  Implications for the Hubble Constant},''
  \href{http://dx.doi.org/10.3847/2041-8213/ab552d}{{\em Astrophys. J. Lett.}
  {\bfseries 886} no.~2, (2019) L27},
  \href{http://arxiv.org/abs/1908.05625}{{\ttfamily arXiv:1908.05625
  [astro-ph.GA]}}.

\bibitem{Wong:2019kwg}
K.~C. Wong {\em et~al.}, ``{H0LiCOW XIII. A 2.4\% measurement of $H_{0}$ from
  lensed quasars: $5.3\sigma$ tension between early and late-Universe
  probes},''
\href{http://arxiv.org/abs/1907.04869}{{\ttfamily arXiv:1907.04869
  [astro-ph.CO]}}.

\bibitem{Birrer:2020tax}
S.~Birrer {\em et~al.}, ``{TDCOSMO IV: Hierarchical time-delay cosmography --
  joint inference of the Hubble constant and galaxy density profiles},''
  \href{http://arxiv.org/abs/2007.02941}{{\ttfamily arXiv:2007.02941
  [astro-ph.CO]}}.

\bibitem{Cardona:2016ems}
W.~Cardona, M.~Kunz, and V.~Pettorino, ``{Determining $H_0$ with Bayesian
  hyper-parameters},''
  \href{http://dx.doi.org/10.1088/1475-7516/2017/03/056}{{\em JCAP} {\bfseries
  1703} no.~03, (2017) 056},
\href{http://arxiv.org/abs/1611.06088}{{\ttfamily arXiv:1611.06088
  [astro-ph.CO]}}.

\bibitem{Camarena:2019moy}
D.~Camarena and V.~Marra, ``{Local determination of the Hubble constant and the
  deceleration parameter},''
  \href{http://dx.doi.org/10.1103/PhysRevResearch.2.013028}{{\em Phys. Rev.
  Res.} {\bfseries 2} no.~1, (2020) 013028},
  \href{http://arxiv.org/abs/1906.11814}{{\ttfamily arXiv:1906.11814
  [astro-ph.CO]}}.

\bibitem{Freedman:2019jwv}
W.~L. Freedman {\em et~al.}, ``{The Carnegie-Chicago Hubble Program. VIII. An
  Independent Determination of the Hubble Constant Based on the Tip of the Red
  Giant Branch},''
\href{http://arxiv.org/abs/1907.05922}{{\ttfamily arXiv:1907.05922
  [astro-ph.CO]}}.

\bibitem{Yuan:2019npk}
W.~Yuan, A.~G. Riess, L.~M. Macri, S.~Casertano, and D.~Scolnic, ``{Consistent
  Calibration of the Tip of the Red Giant Branch in the Large Magellanic Cloud
  on the Hubble Space Telescope Photometric System and a Re-determination of
  the Hubble Constant},''
  \href{http://dx.doi.org/10.3847/1538-4357/ab4bc9}{{\em Astrophys. J.}
  {\bfseries 886} (2019) 61},
\href{http://arxiv.org/abs/1908.00993}{{\ttfamily arXiv:1908.00993
  [astro-ph.GA]}}.

\bibitem{Freedman:2020dne}
W.~L. Freedman, B.~F. Madore, T.~Hoyt, I.~S. Jang, R.~Beaton, M.~G. Lee,
  A.~Monson, J.~Neeley, and J.~Rich, ``{Calibration of the Tip of the Red Giant
  Branch (TRGB)},''
\href{http://arxiv.org/abs/2002.01550}{{\ttfamily arXiv:2002.01550
  [astro-ph.GA]}}.

\bibitem{Khetan:2020hmh}
N.~Khetan {\em et~al.}, ``{A new measurement of the Hubble constant using Type
  Ia supernovae calibrated with surface brightness fluctuations},''
  \href{http://arxiv.org/abs/2008.07754}{{\ttfamily arXiv:2008.07754
  [astro-ph.CO]}}.

\bibitem{Yu:2017iju}
H.~Yu, B.~Ratra, and F.-Y. Wang, ``{Hubble Parameter and Baryon Acoustic
  Oscillation Measurement Constraints on the Hubble Constant, the Deviation
  from the Spatially Flat $\Lambda$CDM Model, the Deceleration--Acceleration
  Transition Redshift, and Spatial Curvature},''
  \href{http://dx.doi.org/10.3847/1538-4357/aab0a2}{{\em Astrophys. J.}
  {\bfseries 856} no.~1, (2018) 3},
  \href{http://arxiv.org/abs/1711.03437}{{\ttfamily arXiv:1711.03437
  [astro-ph.CO]}}.

\bibitem{Gomez-Valent:2019lny}
A.~Gómez-Valent and L.~Amendola, ``{$H_0$ from cosmic chronometers and Type Ia
  supernovae, with Gaussian processes and the weighted polynomial regression
  method},'' in {\em {15th Marcel Grossmann Meeting on Recent Developments in
  Theoretical and Experimental General Relativity, Astrophysics, and
  Relativistic Field Theories}}.
\newblock 5, 2019.
\newblock \href{http://arxiv.org/abs/1905.04052}{{\ttfamily arXiv:1905.04052
  [astro-ph.CO]}}.

\bibitem{Haridasu:2018gqm}
B.~S. Haridasu, V.~V. Lukovi\'c, M.~Moresco, and N.~Vittorio, ``{An improved
  model-independent assessment of the late-time cosmic expansion},''
  \href{http://dx.doi.org/10.1088/1475-7516/2018/10/015}{{\em JCAP} {\bfseries
  10} (2018) 015}, \href{http://arxiv.org/abs/1805.03595}{{\ttfamily
  arXiv:1805.03595 [astro-ph.CO]}}.

\bibitem{Dutta:2019pio}
K.~Dutta, A.~Roy, Ruchika, A.~A. Sen, and M.~Sheikh-Jabbari, ``{Cosmology with
  low-redshift observations: No signal for new physics},''
  \href{http://dx.doi.org/10.1103/PhysRevD.100.103501}{{\em Phys. Rev. D}
  {\bfseries 100} no.~10, (2019) 103501},
  \href{http://arxiv.org/abs/1908.07267}{{\ttfamily arXiv:1908.07267
  [astro-ph.CO]}}.

\bibitem{Huang:2019yhh}
C.~D. Huang, A.~G. Riess, W.~Yuan, L.~M. Macri, N.~L. Zakamska, S.~Casertano,
  P.~A. Whitelock, S.~L. Hoffmann, A.~V. Filippenko, and D.~Scolnic, ``{Hubble
  Space Telescope Observations of Mira Variables in the Type Ia Supernova Host
  NGC 1559: An Alternative Candle to Measure the Hubble Constant},''
  \href{http://arxiv.org/abs/1908.10883}{{\ttfamily arXiv:1908.10883
  [astro-ph.CO]}}.

\bibitem{Shajib:2019toy}
{\bfseries DES} Collaboration, A.~Shajib {\em et~al.}, ``{STRIDES: a 3.9 per
  cent measurement of the Hubble constant from the strong lens system DES
  J0408-5354},'' \href{http://dx.doi.org/10.1093/mnras/staa828}{{\em Mon. Not.
  Roy. Astron. Soc.} {\bfseries 494} no.~4, (2020) 6072--6102},
  \href{http://arxiv.org/abs/1910.06306}{{\ttfamily arXiv:1910.06306
  [astro-ph.CO]}}.

\bibitem{Kourkchi:2020iyz}
E.~Kourkchi, R.~B. Tully, G.~S. Anand, H.~M. Courtois, A.~Dupuy, J.~D. Neill,
  L.~Rizzi, and M.~Seibert, ``{Cosmicflows-4: The Calibration of Optical and
  Infrared Tully--Fisher Relations},''
  \href{http://dx.doi.org/10.3847/1538-4357/ab901c}{{\em Astrophys. J.}
  {\bfseries 896} no.~1, (2020) 3},
  \href{http://arxiv.org/abs/2004.14499}{{\ttfamily arXiv:2004.14499
  [astro-ph.GA]}}.

\bibitem{Schombert:2020pxm}
J.~Schombert, S.~McGaugh, and F.~Lelli, ``{Using The Baryonic Tully-Fisher
  Relation to Measure $H_o$},''
  \href{http://dx.doi.org/10.3847/1538-3881/ab9d88}{{\em Astron. J.} {\bfseries
  160} no.~2, (2020) 71}, \href{http://arxiv.org/abs/2006.08615}{{\ttfamily
  arXiv:2006.08615 [astro-ph.CO]}}.

\bibitem{deJaeger:2020zpb}
T.~de~Jaeger, B.~Stahl, W.~Zheng, A.~Filippenko, A.~Riess, and L.~Galbany, ``{A
  measurement of the Hubble constant from Type II supernovae},''
  \href{http://arxiv.org/abs/2006.03412}{{\ttfamily arXiv:2006.03412
  [astro-ph.CO]}}.

\bibitem{Efstathiou:2020wxn}
G.~Efstathiou, ``{A Lockdown Perspective on the Hubble Tension (with comments
  from the SH0ES team)},'' \href{http://arxiv.org/abs/2007.10716}{{\ttfamily
  arXiv:2007.10716 [astro-ph.CO]}}.

\bibitem{Pesce:2020xfe}
D.~Pesce {\em et~al.}, ``{The Megamaser Cosmology Project. XIII. Combined
  Hubble constant constraints},''
  \href{http://dx.doi.org/10.3847/2041-8213/ab75f0}{{\em Astrophys. J. Lett.}
  {\bfseries 891} no.~1, (2020) L1},
  \href{http://arxiv.org/abs/2001.09213}{{\ttfamily arXiv:2001.09213
  [astro-ph.CO]}}.

\bibitem{Riess:2020sih}
A.~G. Riess, ``{The Expansion of the Universe is Faster than Expected},''
  \href{http://dx.doi.org/10.1038/s42254-019-0137-0}{{\em Nature Rev. Phys.}
  {\bfseries 2} no.~1, (2019) 10--12},
  \href{http://arxiv.org/abs/2001.03624}{{\ttfamily arXiv:2001.03624
  [astro-ph.CO]}}.

\bibitem{Wu:2017fpr}
H.-Y. Wu and D.~Huterer, ``{Sample variance in the local measurements of the
  Hubble constant},'' \href{http://dx.doi.org/10.1093/mnras/stx1967}{{\em Mon.
  Not. Roy. Astron. Soc.} {\bfseries 471} no.~4, (2017) 4946--4955},
  \href{http://arxiv.org/abs/1706.09723}{{\ttfamily arXiv:1706.09723
  [astro-ph.CO]}}.

\bibitem{Kenworthy:2019qwq}
W.~D. Kenworthy, D.~Scolnic, and A.~Riess, ``{The Local Perspective on the
  Hubble Tension: Local Structure Does Not Impact Measurement of the Hubble
  Constant},'' \href{http://dx.doi.org/10.3847/1538-4357/ab0ebf}{{\em
  Astrophys. J.} {\bfseries 875} no.~2, (2019) 145},
  \href{http://arxiv.org/abs/1901.08681}{{\ttfamily arXiv:1901.08681
  [astro-ph.CO]}}.

\bibitem{Tram:2016rcw}
T.~Tram, R.~Vallance, and V.~Vennin, ``{Inflation Model Selection meets Dark
  Radiation},'' \href{http://dx.doi.org/10.1088/1475-7516/2017/01/046}{{\em
  JCAP} {\bfseries 01} (2017) 046},
  \href{http://arxiv.org/abs/1606.09199}{{\ttfamily arXiv:1606.09199
  [astro-ph.CO]}}.

\bibitem{DiValentino:2016ziq}
E.~Di~Valentino and L.~Mersini-Houghton, ``{Testing Predictions of the Quantum
  Landscape Multiverse 2: The Exponential Inflationary Potential},''
  \href{http://dx.doi.org/10.1088/1475-7516/2017/03/020}{{\em JCAP} {\bfseries
  03} (2017) 020}, \href{http://arxiv.org/abs/1612.08334}{{\ttfamily
  arXiv:1612.08334 [astro-ph.CO]}}.

\bibitem{DiValentino:2016hlg}
E.~Di~Valentino, A.~Melchiorri, and J.~Silk, ``{Reconciling Planck with the
  local value of $H_0$ in extended parameter space},''
  \href{http://dx.doi.org/10.1016/j.physletb.2016.08.043}{{\em Phys. Lett.}
  {\bfseries B761} (2016) 242--246},
\href{http://arxiv.org/abs/1606.00634}{{\ttfamily arXiv:1606.00634
  [astro-ph.CO]}}.

\bibitem{Bernal:2016gxb}
J.~L. Bernal, L.~Verde, and A.~G. Riess, ``{The trouble with $H_0$},''
  \href{http://dx.doi.org/10.1088/1475-7516/2016/10/019}{{\em JCAP} {\bfseries
  1610} no.~10, (2016) 019},
\href{http://arxiv.org/abs/1607.05617}{{\ttfamily arXiv:1607.05617
  [astro-ph.CO]}}.

\bibitem{DiValentino:2017oaw}
E.~Di~Valentino, C.~Bøehm, E.~Hivon, and F.~R. Bouchet, ``{Reducing the $H_0$
  and $\sigma_8$ tensions with Dark Matter-neutrino interactions},''
  \href{http://dx.doi.org/10.1103/PhysRevD.97.043513}{{\em Phys. Rev.}
  {\bfseries D97} no.~4, (2018) 043513},
\href{http://arxiv.org/abs/1710.02559}{{\ttfamily arXiv:1710.02559
  [astro-ph.CO]}}.

\bibitem{DiValentino:2017rcr}
E.~Di~Valentino, E.~V. Linder, and A.~Melchiorri, ``{Vacuum phase transition
  solves the $H_0$ tension},''
  \href{http://dx.doi.org/10.1103/PhysRevD.97.043528}{{\em Phys. Rev.}
  {\bfseries D97} no.~4, (2018) 043528},
\href{http://arxiv.org/abs/1710.02153}{{\ttfamily arXiv:1710.02153
  [astro-ph.CO]}}.

\bibitem{Binder:2017lkj}
T.~Binder, M.~Gustafsson, A.~Kamada, S.~M.~R. Sandner, and M.~Wiesner,
  ``{Reannihilation of self-interacting dark matter},''
  \href{http://dx.doi.org/10.1103/PhysRevD.97.123004}{{\em Phys. Rev.}
  {\bfseries D97} no.~12, (2018) 123004},
\href{http://arxiv.org/abs/1712.01246}{{\ttfamily arXiv:1712.01246
  [astro-ph.CO]}}.

\bibitem{Khosravi:2017hfi}
N.~Khosravi, S.~Baghram, N.~Afshordi, and N.~Altamirano, ``{$H_0$ tension as a
  hint for a transition in gravitational theory},''
  \href{http://dx.doi.org/10.1103/PhysRevD.99.103526}{{\em Phys. Rev.}
  {\bfseries D99} no.~10, (2019) 103526},
\href{http://arxiv.org/abs/1710.09366}{{\ttfamily arXiv:1710.09366
  [astro-ph.CO]}}.

\bibitem{DiValentino:2017zyq}
E.~Di~Valentino, A.~Melchiorri, E.~V. Linder, and J.~Silk, ``{Constraining Dark
  Energy Dynamics in Extended Parameter Space},''
  \href{http://dx.doi.org/10.1103/PhysRevD.96.023523}{{\em Phys. Rev.}
  {\bfseries D96} no.~2, (2017) 023523},
\href{http://arxiv.org/abs/1704.00762}{{\ttfamily arXiv:1704.00762
  [astro-ph.CO]}}.

\bibitem{Renk:2017rzu}
J.~Renk, M.~Zumalacárregui, F.~Montanari, and A.~Barreira, ``{Galileon gravity
  in light of ISW, CMB, BAO and H$_0$ data},''
  \href{http://dx.doi.org/10.1088/1475-7516/2017/10/020}{{\em JCAP} {\bfseries
  1710} no.~10, (2017) 020},
\href{http://arxiv.org/abs/1707.02263}{{\ttfamily arXiv:1707.02263
  [astro-ph.CO]}}.

\bibitem{DiValentino:2017gzb}
E.~Di~Valentino, ``{Crack in the cosmological paradigm},''
  \href{http://dx.doi.org/10.1038/s41550-017-0236-8}{{\em Nat. Astron.}
  {\bfseries 1} no.~9, (2017) 569--570},
\href{http://arxiv.org/abs/1709.04046}{{\ttfamily arXiv:1709.04046
  [physics.pop-ph]}}.

\bibitem{Kumar:2016zpg}
S.~Kumar and R.~C. Nunes, ``{Probing the interaction between dark matter and
  dark energy in the presence of massive neutrinos},''
  \href{http://dx.doi.org/10.1103/PhysRevD.94.123511}{{\em Phys. Rev.}
  {\bfseries D94} no.~12, (2016) 123511},
\href{http://arxiv.org/abs/1608.02454}{{\ttfamily arXiv:1608.02454
  [astro-ph.CO]}}.

\bibitem{DiValentino:2017iww}
E.~Di~Valentino, A.~Melchiorri, and O.~Mena, ``{Can interacting dark energy
  solve the $H_0$ tension?},''
  \href{http://dx.doi.org/10.1103/PhysRevD.96.043503}{{\em Phys. Rev.}
  {\bfseries D96} no.~4, (2017) 043503},
\href{http://arxiv.org/abs/1704.08342}{{\ttfamily arXiv:1704.08342
  [astro-ph.CO]}}.

\bibitem{Kumar:2017dnp}
S.~Kumar and R.~C. Nunes, ``{Echo of interactions in the dark sector},''
  \href{http://dx.doi.org/10.1103/PhysRevD.96.103511}{{\em Phys. Rev.}
  {\bfseries D96} no.~10, (2017) 103511},
\href{http://arxiv.org/abs/1702.02143}{{\ttfamily arXiv:1702.02143
  [astro-ph.CO]}}.

\bibitem{Gomez-Valent:2020mqn}
A.~Gómez-Valent, V.~Pettorino, and L.~Amendola, ``{Update on coupled dark
  energy and the $H_0$ tension},''
  \href{http://dx.doi.org/10.1103/PhysRevD.101.123513}{{\em Phys. Rev. D}
  {\bfseries 101} no.~12, (2020) 123513},
  \href{http://arxiv.org/abs/2004.00610}{{\ttfamily arXiv:2004.00610
  [astro-ph.CO]}}.

\bibitem{Lucca:2020zjb}
M.~Lucca and D.~C. Hooper, ``{Tensions in the dark: shedding light on Dark
  Matter-Dark Energy interactions},''
  \href{http://arxiv.org/abs/2002.06127}{{\ttfamily arXiv:2002.06127
  [astro-ph.CO]}}.

\bibitem{vandeBruck:2017idm}
C.~Van De~Bruck and J.~Mifsud, ``{Searching for dark matter - dark energy
  interactions: going beyond the conformal case},''
  \href{http://dx.doi.org/10.1103/PhysRevD.97.023506}{{\em Phys. Rev. D}
  {\bfseries 97} no.~2, (2018) 023506},
  \href{http://arxiv.org/abs/1709.04882}{{\ttfamily arXiv:1709.04882
  [astro-ph.CO]}}.

\bibitem{Yang:2018euj}
W.~Yang, S.~Pan, E.~Di~Valentino, R.~C. Nunes, S.~Vagnozzi, and D.~F. Mota,
  ``{Tale of stable interacting dark energy, observational signatures, and the
  $H_0$ tension},'' \href{http://dx.doi.org/10.1088/1475-7516/2018/09/019}{{\em
  JCAP} {\bfseries 1809} no.~09, (2018) 019},
\href{http://arxiv.org/abs/1805.08252}{{\ttfamily arXiv:1805.08252
  [astro-ph.CO]}}.

\bibitem{Yang:2018uae}
W.~Yang, A.~Mukherjee, E.~Di~Valentino, and S.~Pan, ``{Interacting dark energy
  with time varying equation of state and the $H_0$ tension},''
  \href{http://dx.doi.org/10.1103/PhysRevD.98.123527}{{\em Phys. Rev.}
  {\bfseries D98} no.~12, (2018) 123527},
\href{http://arxiv.org/abs/1809.06883}{{\ttfamily arXiv:1809.06883
  [astro-ph.CO]}}.

\bibitem{Yang:2019uzo}
W.~Yang, O.~Mena, S.~Pan, and E.~Di~Valentino, ``{Dark sectors with dynamical
  coupling},'' \href{http://dx.doi.org/10.1103/PhysRevD.100.083509}{{\em Phys.
  Rev.} {\bfseries D100} no.~8, (2019) 083509},
\href{http://arxiv.org/abs/1906.11697}{{\ttfamily arXiv:1906.11697
  [astro-ph.CO]}}.

\bibitem{Martinelli:2019dau}
M.~Martinelli, N.~B. Hogg, S.~Peirone, M.~Bruni, and D.~Wands, ``{Constraints
  on the interacting vacuum–geodesic CDM scenario},''
  \href{http://dx.doi.org/10.1093/mnras/stz1915}{{\em Mon. Not. Roy. Astron.
  Soc.} {\bfseries 488} no.~3, (2019) 3423--3438},
\href{http://arxiv.org/abs/1902.10694}{{\ttfamily arXiv:1902.10694
  [astro-ph.CO]}}.

\bibitem{DiValentino:2019ffd}
E.~Di~Valentino, A.~Melchiorri, O.~Mena, and S.~Vagnozzi, ``{Interacting dark
  energy in the early 2020s: a promising solution to the $H_0$ and cosmic shear
  tensions},'' \href{http://dx.doi.org/10.1016/j.dark.2020.100666}{{\em Phys.
  Dark Univ.} {\bfseries 30} (2020) 100666},
  \href{http://arxiv.org/abs/1908.04281}{{\ttfamily arXiv:1908.04281
  [astro-ph.CO]}}.

\bibitem{DiValentino:2019jae}
E.~Di~Valentino, A.~Melchiorri, O.~Mena, and S.~Vagnozzi, ``{Nonminimal dark
  sector physics and cosmological tensions},''
  \href{http://dx.doi.org/10.1103/PhysRevD.101.063502}{{\em Phys. Rev. D}
  {\bfseries 101} no.~6, (2020) 063502},
  \href{http://arxiv.org/abs/1910.09853}{{\ttfamily arXiv:1910.09853
  [astro-ph.CO]}}.

\bibitem{Benevento:2020fev}
G.~Benevento, W.~Hu, and M.~Raveri, ``{Can Late Dark Energy Transitions Raise
  the Hubble constant?},''
\href{http://arxiv.org/abs/2002.11707}{{\ttfamily arXiv:2002.11707
  [astro-ph.CO]}}.

\bibitem{Belgacem:2017cqo}
E.~Belgacem, Y.~Dirian, S.~Foffa, and M.~Maggiore, ``{Nonlocal gravity.
  Conceptual aspects and cosmological predictions},''
  \href{http://dx.doi.org/10.1088/1475-7516/2018/03/002}{{\em JCAP} {\bfseries
  1803} no.~03, (2018) 002},
\href{http://arxiv.org/abs/1712.07066}{{\ttfamily arXiv:1712.07066 [hep-th]}}.

\bibitem{Fernandez-Arenas:2017isq}
D.~Fernández~Arenas, E.~Terlevich, R.~Terlevich, J.~Melnick, R.~Chávez,
  F.~Bresolin, E.~Telles, M.~Plionis, and S.~Basilakos, ``{An independent
  determination of the local Hubble constant},''
  \href{http://dx.doi.org/10.1093/mnras/stx2710}{{\em Mon. Not. Roy. Astron.
  Soc.} {\bfseries 474} no.~1, (2018) 1250--1276},
\href{http://arxiv.org/abs/1710.05951}{{\ttfamily arXiv:1710.05951
  [astro-ph.CO]}}.

\bibitem{Sola:2017znb}
J.~Solà, A.~Gómez-Valent, and J.~de~Cruz~Pérez, ``{The $H_0$ tension in
  light of vacuum dynamics in the Universe},''
  \href{http://dx.doi.org/10.1016/j.physletb.2017.09.073}{{\em Phys. Lett.}
  {\bfseries B774} (2017) 317--324},
\href{http://arxiv.org/abs/1705.06723}{{\ttfamily arXiv:1705.06723
  [astro-ph.CO]}}.

\bibitem{Nunes:2018xbm}
R.~C. Nunes, ``{Structure formation in $f(T)$ gravity and a solution for $H_0$
  tension},'' \href{http://dx.doi.org/10.1088/1475-7516/2018/05/052}{{\em JCAP}
  {\bfseries 1805} no.~05, (2018) 052},
\href{http://arxiv.org/abs/1802.02281}{{\ttfamily arXiv:1802.02281 [gr-qc]}}.

\bibitem{Colgain:2018wgk}
E.~Ó~Colgáin, M.~H. P.~M. van Putten, and H.~Yavartanoo, ``{de Sitter
  Swampland, $H_0$ tension \& observation},''
  \href{http://dx.doi.org/10.1016/j.physletb.2019.04.032}{{\em Phys. Lett.}
  {\bfseries B793} (2019) 126--129},
\href{http://arxiv.org/abs/1807.07451}{{\ttfamily arXiv:1807.07451 [hep-th]}}.

\bibitem{DEramo:2018vss}
F.~D'Eramo, R.~Z. Ferreira, A.~Notari, and J.~L. Bernal, ``{Hot Axions and the
  $H_0$ tension},'' \href{http://dx.doi.org/10.1088/1475-7516/2018/11/014}{{\em
  JCAP} {\bfseries 1811} no.~11, (2018) 014},
\href{http://arxiv.org/abs/1808.07430}{{\ttfamily arXiv:1808.07430 [hep-ph]}}.

\bibitem{Guo:2018ans}
R.-Y. Guo, J.-F. Zhang, and X.~Zhang, ``{Can the $H_0$ tension be resolved in
  extensions to $\Lambda$CDM cosmology?},''
  \href{http://dx.doi.org/10.1088/1475-7516/2019/02/054}{{\em JCAP} {\bfseries
  1902} (2019) 054},
\href{http://arxiv.org/abs/1809.02340}{{\ttfamily arXiv:1809.02340
  [astro-ph.CO]}}.

\bibitem{Lin:2018nxe}
M.-X. Lin, M.~Raveri, and W.~Hu, ``{Phenomenology of Modified Gravity at
  Recombination},'' \href{http://dx.doi.org/10.1103/PhysRevD.99.043514}{{\em
  Phys. Rev.} {\bfseries D99} no.~4, (2019) 043514},
\href{http://arxiv.org/abs/1810.02333}{{\ttfamily arXiv:1810.02333
  [astro-ph.CO]}}.

\bibitem{Yang:2018qmz}
W.~Yang, S.~Pan, E.~Di~Valentino, E.~N. Saridakis, and S.~Chakraborty,
  ``{Observational constraints on one-parameter dynamical dark-energy
  parametrizations and the $H_0$ tension},''
  \href{http://dx.doi.org/10.1103/PhysRevD.99.043543}{{\em Phys. Rev.}
  {\bfseries D99} no.~4, (2019) 043543},
\href{http://arxiv.org/abs/1810.05141}{{\ttfamily arXiv:1810.05141
  [astro-ph.CO]}}.

\bibitem{Vagnozzi:2019ezj}
S.~Vagnozzi, ``{New physics in light of the $H_0$ tension: an alternative
  view},'' \href{http://dx.doi.org/10.1103/PhysRevD.102.023518}{{\em Phys. Rev.
  D} {\bfseries 102} no.~2, (2020) 023518},
  \href{http://arxiv.org/abs/1907.07569}{{\ttfamily arXiv:1907.07569
  [astro-ph.CO]}}.

\bibitem{DiValentino:2019dzu}
E.~Di~Valentino, A.~Melchiorri, and J.~Silk, ``{Cosmological constraints in
  extended parameter space from the Planck 2018 Legacy release},''
  \href{http://dx.doi.org/10.1088/1475-7516/2020/01/013}{{\em JCAP} {\bfseries
  01} (2020) 013}, \href{http://arxiv.org/abs/1908.01391}{{\ttfamily
  arXiv:1908.01391 [astro-ph.CO]}}.

\bibitem{DiValentino:2020naf}
E.~Di~Valentino, A.~Mukherjee, and A.~A. Sen, ``{Dark Energy with Phantom
  Crossing and the $H_0$ tension},''
  \href{http://arxiv.org/abs/2005.12587}{{\ttfamily arXiv:2005.12587
  [astro-ph.CO]}}.

\bibitem{Keeley:2019esp}
R.~E. Keeley, S.~Joudaki, M.~Kaplinghat, and D.~Kirkby, ``{Implications of a
  transition in the dark energy equation of state for the $H_0$ and $\sigma_8$
  tensions},'' \href{http://dx.doi.org/10.1088/1475-7516/2019/12/035}{{\em
  JCAP} {\bfseries 12} (2019) 035},
  \href{http://arxiv.org/abs/1905.10198}{{\ttfamily arXiv:1905.10198
  [astro-ph.CO]}}.

\bibitem{Joudaki:2016kym}
S.~Joudaki {\em et~al.}, ``{KiDS-450: Testing extensions to the standard
  cosmological model},'' \href{http://dx.doi.org/10.1093/mnras/stx998}{{\em
  Mon. Not. Roy. Astron. Soc.} {\bfseries 471} no.~2, (2017) 1259--1279},
  \href{http://arxiv.org/abs/1610.04606}{{\ttfamily arXiv:1610.04606
  [astro-ph.CO]}}.

\bibitem{Yang:2018prh}
W.~Yang, S.~Pan, E.~Di~Valentino, and E.~N. Saridakis, ``{Observational
  constraints on dynamical dark energy with pivoting redshift},''
  \href{http://dx.doi.org/10.3390/universe5110219}{{\em Universe} {\bfseries 5}
  no.~11, (2019) 219}, \href{http://arxiv.org/abs/1811.06932}{{\ttfamily
  arXiv:1811.06932 [astro-ph.CO]}}.

\bibitem{Poulin:2018cxd}
V.~Poulin, T.~L. Smith, T.~Karwal, and M.~Kamionkowski, ``{Early Dark Energy
  Can Resolve The Hubble Tension},''
  \href{http://dx.doi.org/10.1103/PhysRevLett.122.221301}{{\em Phys. Rev.
  Lett.} {\bfseries 122} no.~22, (2019) 221301},
\href{http://arxiv.org/abs/1811.04083}{{\ttfamily arXiv:1811.04083
  [astro-ph.CO]}}.

\bibitem{Karwal:2016vyq}
T.~Karwal and M.~Kamionkowski, ``{Dark energy at early times, the Hubble
  parameter, and the string axiverse},''
  \href{http://dx.doi.org/10.1103/PhysRevD.94.103523}{{\em Phys. Rev. D}
  {\bfseries 94} no.~10, (2016) 103523},
  \href{http://arxiv.org/abs/1608.01309}{{\ttfamily arXiv:1608.01309
  [astro-ph.CO]}}.

\bibitem{Banihashemi:2018oxo}
A.~Banihashemi, N.~Khosravi, and A.~H. Shirazi, ``{Phase transition in the dark
  sector as a proposal to lessen cosmological tensions},''
  \href{http://dx.doi.org/10.1103/PhysRevD.101.123521}{{\em Phys. Rev. D}
  {\bfseries 101} no.~12, (2020) 123521},
  \href{http://arxiv.org/abs/1808.02472}{{\ttfamily arXiv:1808.02472
  [astro-ph.CO]}}.

\bibitem{Banihashemi:2018has}
A.~Banihashemi, N.~Khosravi, and A.~H. Shirazi, ``{Ginzburg-Landau Theory of
  Dark Energy: A Framework to Study Both Temporal and Spatial Cosmological
  Tensions Simultaneously},''
  \href{http://dx.doi.org/10.1103/PhysRevD.99.083509}{{\em Phys. Rev.}
  {\bfseries D99} no.~8, (2019) 083509},
\href{http://arxiv.org/abs/1810.11007}{{\ttfamily arXiv:1810.11007
  [astro-ph.CO]}}.

\bibitem{Mortsell:2018mfj}
E.~Mörtsell and S.~Dhawan, ``{Does the Hubble constant tension call for new
  physics?},'' \href{http://dx.doi.org/10.1088/1475-7516/2018/09/025}{{\em
  JCAP} {\bfseries 1809} no.~09, (2018) 025},
\href{http://arxiv.org/abs/1801.07260}{{\ttfamily arXiv:1801.07260
  [astro-ph.CO]}}.

\bibitem{Zhang:2018air}
X.~Zhang and Q.-G. Huang, ``{Constraints on $H_0$ from WMAP and BAO
  Measurements},'' \href{http://dx.doi.org/10.1088/0253-6102/71/7/826}{{\em
  Commun. Theor. Phys.} {\bfseries 71} no.~7, (2019) 826--830},
\href{http://arxiv.org/abs/1812.01877}{{\ttfamily arXiv:1812.01877
  [astro-ph.CO]}}.

\bibitem{Kreisch:2019yzn}
C.~D. Kreisch, F.-Y. Cyr-Racine, and O.~Doré, ``{The Neutrino Puzzle:
  Anomalies, Interactions, and Cosmological Tensions},''
  \href{http://dx.doi.org/10.1103/PhysRevD.101.123505}{{\em Phys. Rev. D}
  {\bfseries 101} no.~12, (2020) 123505},
  \href{http://arxiv.org/abs/1902.00534}{{\ttfamily arXiv:1902.00534
  [astro-ph.CO]}}.

\bibitem{Kumar:2019wfs}
S.~Kumar, R.~C. Nunes, and S.~K. Yadav, ``{Dark sector interaction: a remedy of
  the tensions between CMB and LSS data},''
  \href{http://dx.doi.org/10.1140/epjc/s10052-019-7087-7}{{\em Eur. Phys. J.}
  {\bfseries C79} no.~7, (2019) 576},
\href{http://arxiv.org/abs/1903.04865}{{\ttfamily arXiv:1903.04865
  [astro-ph.CO]}}.

\bibitem{Agrawal:2019lmo}
P.~Agrawal, F.-Y. Cyr-Racine, D.~Pinner, and L.~Randall, ``{Rock 'n' Roll
  Solutions to the Hubble Tension},''
\href{http://arxiv.org/abs/1904.01016}{{\ttfamily arXiv:1904.01016
  [astro-ph.CO]}}.

\bibitem{Yang:2019jwn}
W.~Yang, S.~Pan, A.~Paliathanasis, S.~Ghosh, and Y.~Wu, ``{Observational
  constraints of a new unified dark fluid and the $H_0$ tension},''
  \href{http://dx.doi.org/10.1093/mnras/stz2753}{{\em Mon. Not. Roy. Astron.
  Soc.} {\bfseries 490} no.~2, (2019) 2071--2085},
\href{http://arxiv.org/abs/1904.10436}{{\ttfamily arXiv:1904.10436 [gr-qc]}}.

\bibitem{Yang:2019qza}
W.~Yang, S.~Pan, E.~Di~Valentino, A.~Paliathanasis, and J.~Lu, ``{Challenging
  bulk viscous unified scenarios with cosmological observations},''
  \href{http://dx.doi.org/10.1103/PhysRevD.100.103518}{{\em Phys. Rev.}
  {\bfseries D100} no.~10, (2019) 103518},
\href{http://arxiv.org/abs/1906.04162}{{\ttfamily arXiv:1906.04162
  [astro-ph.CO]}}.

\bibitem{DiValentino:2019exe}
E.~Di~Valentino, R.~Z. Ferreira, L.~Visinelli, and U.~Danielsson, ``{Late time
  transitions in the quintessence field and the $H_0$ tension},''
  \href{http://dx.doi.org/10.1016/j.dark.2019.100385}{{\em Phys. Dark Univ.}
  {\bfseries 26} (2019) 100385},
\href{http://arxiv.org/abs/1906.11255}{{\ttfamily arXiv:1906.11255
  [astro-ph.CO]}}.

\bibitem{Desmond:2019ygn}
H.~Desmond, B.~Jain, and J.~Sakstein, ``{Local resolution of the Hubble
  tension: The impact of screened fifth forces on the cosmic distance
  ladder},'' \href{http://dx.doi.org/10.1103/PhysRevD.100.043537}{{\em Phys.
  Rev.} {\bfseries D100} no.~4, (2019) 043537},
\href{http://arxiv.org/abs/1907.03778}{{\ttfamily arXiv:1907.03778
  [astro-ph.CO]}}.

\bibitem{Yang:2019nhz}
W.~Yang, S.~Pan, S.~Vagnozzi, E.~Di~Valentino, D.~F. Mota, and S.~Capozziello,
  ``{Dawn of the dark: unified dark sectors and the EDGES Cosmic Dawn 21-cm
  signal},'' \href{http://dx.doi.org/10.1088/1475-7516/2019/11/044}{{\em JCAP}
  {\bfseries 1911} (2019) 044},
\href{http://arxiv.org/abs/1907.05344}{{\ttfamily arXiv:1907.05344
  [astro-ph.CO]}}.

\bibitem{Pan:2019gop}
S.~Pan, W.~Yang, E.~Di~Valentino, E.~N. Saridakis, and S.~Chakraborty,
  ``{Interacting scenarios with dynamical dark energy: Observational
  constraints and alleviation of the $H_0$ tension},''
  \href{http://dx.doi.org/10.1103/PhysRevD.100.103520}{{\em Phys. Rev.}
  {\bfseries D100} no.~10, (2019) 103520},
\href{http://arxiv.org/abs/1907.07540}{{\ttfamily arXiv:1907.07540
  [astro-ph.CO]}}.

\bibitem{Visinelli:2019qqu}
L.~Visinelli, S.~Vagnozzi, and U.~Danielsson, ``{Revisiting a negative
  cosmological constant from low-redshift data},''
  \href{http://dx.doi.org/10.3390/sym11081035}{{\em Symmetry} {\bfseries 11}
  no.~8, (2019) 1035},
\href{http://arxiv.org/abs/1907.07953}{{\ttfamily arXiv:1907.07953
  [astro-ph.CO]}}.

\bibitem{Smith:2019ihp}
T.~L. Smith, V.~Poulin, and M.~A. Amin, ``{Oscillating scalar fields and the
  Hubble tension: a resolution with novel signatures},''
  \href{http://dx.doi.org/10.1103/PhysRevD.101.063523}{{\em Phys. Rev. D}
  {\bfseries 101} no.~6, (2020) 063523},
  \href{http://arxiv.org/abs/1908.06995}{{\ttfamily arXiv:1908.06995
  [astro-ph.CO]}}.

\bibitem{Lucca:2020fgp}
M.~Lucca, ``{The role of CMB spectral distortions in the Hubble tension: a
  proof of principle},'' \href{http://arxiv.org/abs/2008.01115}{{\ttfamily
  arXiv:2008.01115 [astro-ph.CO]}}.

\bibitem{Lin:2019qug}
M.-X. Lin, G.~Benevento, W.~Hu, and M.~Raveri, ``{Acoustic Dark Energy:
  Potential Conversion of the Hubble Tension},''
  \href{http://dx.doi.org/10.1103/PhysRevD.100.063542}{{\em Phys. Rev. D}
  {\bfseries 100} no.~6, (2019) 063542},
  \href{http://arxiv.org/abs/1905.12618}{{\ttfamily arXiv:1905.12618
  [astro-ph.CO]}}.

\bibitem{Martinelli:2019krf}
M.~Martinelli and I.~Tutusaus, ``{CMB tensions with low-redshift $H_0$ and
  $S_8$ measurements: impact of a redshift-dependent type-Ia supernovae
  intrinsic luminosity},'' \href{http://dx.doi.org/10.3390/sym11080986}{{\em
  Symmetry} {\bfseries 11} no.~8, (2019) 986},
\href{http://arxiv.org/abs/1906.09189}{{\ttfamily arXiv:1906.09189
  [astro-ph.CO]}}.

\bibitem{Cai:2019bdh}
Y.-F. Cai, M.~Khurshudyan, and E.~N. Saridakis, ``{Model-independent
  reconstruction of $f(T)$ gravity from Gaussian Processes},''
  \href{http://dx.doi.org/10.3847/1538-4357/ab5a7f}{{\em Astrophys. J.}
  {\bfseries 888} (2020) 62}, \href{http://arxiv.org/abs/1907.10813}{{\ttfamily
  arXiv:1907.10813 [astro-ph.CO]}}.

\bibitem{Schoneberg:2019wmt}
N.~Schöneberg, J.~Lesgourgues, and D.~C. Hooper, ``{The BAO+BBN take on the
  Hubble tension},''
  \href{http://dx.doi.org/10.1088/1475-7516/2019/10/029}{{\em JCAP} {\bfseries
  1910} no.~10, (2019) 029},
\href{http://arxiv.org/abs/1907.11594}{{\ttfamily arXiv:1907.11594
  [astro-ph.CO]}}.

\bibitem{Shafieloo:2016bpk}
A.~Shafieloo, D.~K. Hazra, V.~Sahni, and A.~A. Starobinsky, ``{Metastable Dark
  Energy with Radioactive-like Decay},''
  \href{http://dx.doi.org/10.1093/mnras/stx2481}{{\em Mon. Not. Roy. Astron.
  Soc.} {\bfseries 473} no.~2, (2018) 2760--2770},
\href{http://arxiv.org/abs/1610.05192}{{\ttfamily arXiv:1610.05192
  [astro-ph.CO]}}.

\bibitem{Li:2019san}
X.~Li, A.~Shafieloo, V.~Sahni, and A.~A. Starobinsky, ``{Revisiting Metastable
  Dark Energy and Tensions in the Estimation of Cosmological Parameters},''
  \href{http://dx.doi.org/10.3847/1538-4357/ab535d}{{\em Astrophys. J.}
  {\bfseries 887} (4, 2019) 153},
  \href{http://arxiv.org/abs/1904.03790}{{\ttfamily arXiv:1904.03790
  [astro-ph.CO]}}.

\bibitem{Cuceu:2019for}
A.~Cuceu, J.~Farr, P.~Lemos, and A.~Font-Ribera, ``{Baryon Acoustic
  Oscillations and the Hubble Constant: Past, Present and Future},''
  \href{http://dx.doi.org/10.1088/1475-7516/2019/10/044}{{\em JCAP} {\bfseries
  1910} no.~10, (2019) 044},
\href{http://arxiv.org/abs/1906.11628}{{\ttfamily arXiv:1906.11628
  [astro-ph.CO]}}.

\bibitem{Colgain:2019joh}
E.~. Colgáin and H.~Yavartanoo, ``{Testing the Swampland: $H_0$ tension},''
  \href{http://dx.doi.org/10.1016/j.physletb.2019.134907}{{\em Phys. Lett.}
  {\bfseries B797} (2019) 134907},
\href{http://arxiv.org/abs/1905.02555}{{\ttfamily arXiv:1905.02555
  [astro-ph.CO]}}.

\bibitem{Pan:2019jqh}
S.~Pan, W.~Yang, C.~Singha, and E.~N. Saridakis, ``{Observational constraints
  on sign-changeable interaction models and alleviation of the $H_0$
  tension},'' \href{http://dx.doi.org/10.1103/PhysRevD.100.083539}{{\em Phys.
  Rev.} {\bfseries D100} no.~8, (2019) 083539},
\href{http://arxiv.org/abs/1903.10969}{{\ttfamily arXiv:1903.10969
  [astro-ph.CO]}}.

\bibitem{Berghaus:2019cls}
K.~V. Berghaus and T.~Karwal, ``{Thermal Friction as a Solution to the Hubble
  Tension},'' \href{http://dx.doi.org/10.1103/PhysRevD.101.083537}{{\em Phys.
  Rev. D} {\bfseries 101} no.~8, (2020) 083537},
  \href{http://arxiv.org/abs/1911.06281}{{\ttfamily arXiv:1911.06281
  [astro-ph.CO]}}.

\bibitem{Knox:2019rjx}
L.~Knox and M.~Millea, ``{Hubble constant hunter's guide},''
  \href{http://dx.doi.org/10.1103/PhysRevD.101.043533}{{\em Phys. Rev. D}
  {\bfseries 101} no.~4, (2020) 043533},
  \href{http://arxiv.org/abs/1908.03663}{{\ttfamily arXiv:1908.03663
  [astro-ph.CO]}}.

\bibitem{Pandey:2019plg}
K.~L. Pandey, T.~Karwal, and S.~Das, ``{Alleviating the $H_0$ and $\sigma_8$
  anomalies with a decaying dark matter model},''
  \href{http://dx.doi.org/10.1088/1475-7516/2020/07/026}{{\em JCAP} {\bfseries
  07} (2020) 026}, \href{http://arxiv.org/abs/1902.10636}{{\ttfamily
  arXiv:1902.10636 [astro-ph.CO]}}.

\bibitem{Adhikari:2019fvb}
S.~Adhikari and D.~Huterer, ``{Super-CMB fluctuations and the Hubble
  tension},'' \href{http://dx.doi.org/10.1016/j.dark.2020.100539}{{\em Phys.
  Dark Univ.} {\bfseries 28} (2020) 100539},
  \href{http://arxiv.org/abs/1905.02278}{{\ttfamily arXiv:1905.02278
  [astro-ph.CO]}}.

\bibitem{Lancaster:2017ksf}
L.~Lancaster, F.-Y. Cyr-Racine, L.~Knox, and Z.~Pan, ``{A tale of two modes:
  Neutrino free-streaming in the early universe},''
  \href{http://dx.doi.org/10.1088/1475-7516/2017/07/033}{{\em JCAP} {\bfseries
  1707} no.~07, (2017) 033},
\href{http://arxiv.org/abs/1704.06657}{{\ttfamily arXiv:1704.06657
  [astro-ph.CO]}}.

\bibitem{Niedermann:2019olb}
F.~Niedermann and M.~S. Sloth, ``{New Early Dark Energy},''
\href{http://arxiv.org/abs/1910.10739}{{\ttfamily arXiv:1910.10739
  [astro-ph.CO]}}.

\bibitem{Yadav:2019jio}
S.~K. Yadav, ``{Constraints on Dark Matter-Photon Coupling in the Presence of
  Time-Varying Dark Energy},''
  \href{http://dx.doi.org/10.1142/S0217732319503589}{{\em Mod. Phys. Lett.}
  {\bfseries A33} (2019) 1950358},
\href{http://arxiv.org/abs/1907.05886}{{\ttfamily arXiv:1907.05886
  [astro-ph.CO]}}.

\bibitem{Kasai:2019yqn}
M.~Kasai and T.~Futamase, ``{A possible solution to the Hubble constant
  discrepancy -- Cosmology where the local volume expansion is driven by the
  domain average density},'' \href{http://dx.doi.org/10.1093/ptep/ptz066}{{\em
  PTEP} {\bfseries 2019} no.~7, (2019) 073E01},
\href{http://arxiv.org/abs/1904.09689}{{\ttfamily arXiv:1904.09689 [gr-qc]}}.

\bibitem{Amirhashchi:2020qep}
H.~Amirhashchi and A.~K. Yadav, ``{Interacting Dark Sectors in Anisotropic
  Universe: Observational Constraints and $H_{0}$ Tension},''
\href{http://arxiv.org/abs/2001.03775}{{\ttfamily arXiv:2001.03775
  [astro-ph.CO]}}.

\bibitem{Perez:2020cwa}
A.~Perez, D.~Sudarsky, and E.~Wilson-Ewing, ``{Resolving the $H_0$ tension with
  diffusion},''
\href{http://arxiv.org/abs/2001.07536}{{\ttfamily arXiv:2001.07536
  [astro-ph.CO]}}.

\bibitem{Pan:2020bur}
S.~Pan, W.~Yang, and A.~Paliathanasis, ``{Non-linear interacting cosmological
  models after Planck 2018 legacy release and the $H_0$ tension},''
  \href{http://dx.doi.org/10.1093/mnras/staa213}{{\em Mon. Not. Roy. Astron.
  Soc.} {\bfseries 493} no.~3, (2020) 3114--3131},
  \href{http://arxiv.org/abs/2002.03408}{{\ttfamily arXiv:2002.03408
  [astro-ph.CO]}}.

\bibitem{DAgostino:2020dhv}
R.~D'Agostino and R.~C. Nunes, ``{Measurements of $H_0$ in modified gravity
  theories},''
\href{http://arxiv.org/abs/2002.06381}{{\ttfamily arXiv:2002.06381
  [astro-ph.CO]}}.

\bibitem{Liao:2020zko}
K.~Liao, A.~Shafieloo, R.~E. Keeley, and E.~V. Linder, ``{Determining $H_0$
  Model-Independently and Consistency Tests},''
\href{http://arxiv.org/abs/2002.10605}{{\ttfamily arXiv:2002.10605
  [astro-ph.CO]}}.

\bibitem{Yang:2020zuk}
W.~Yang, E.~Di~Valentino, S.~Pan, S.~Basilakos, and A.~Paliathanasis,
  ``{Metastable dark energy models in light of Planck 2018: Alleviating the
  $H_0$ tension},''
\href{http://arxiv.org/abs/2001.04307}{{\ttfamily arXiv:2001.04307
  [astro-ph.CO]}}.

\bibitem{Pan:2020zza}
S.~Pan, G.~S. Sharov, and W.~Yang, ``{Field theoretic interpretations of
  interacting dark energy scenarios and recent observations},''
  \href{http://dx.doi.org/10.1103/PhysRevD.101.103533}{{\em Phys. Rev. D}
  {\bfseries 101} no.~10, (2020) 103533},
  \href{http://arxiv.org/abs/2001.03120}{{\ttfamily arXiv:2001.03120
  [astro-ph.CO]}}.

\bibitem{Wu:2020nxz}
W.~K. Wu, P.~Motloch, W.~Hu, and M.~Raveri, ``{Hubble constant tension between
  CMB lensing and BAO measurements},''
  \href{http://arxiv.org/abs/2004.10207}{{\ttfamily arXiv:2004.10207
  [astro-ph.CO]}}.

\bibitem{Blinov:2020hmc}
N.~Blinov and G.~Marques-Tavares, ``{Interacting radiation after Planck and its
  implications for the Hubble Tension},''
  \href{http://arxiv.org/abs/2003.08387}{{\ttfamily arXiv:2003.08387
  [astro-ph.CO]}}.

\bibitem{Wang:2020zfv}
D.~Wang and D.~Mota, ``{Can $f(T)$ gravity resolve the $H_0$ tension?},''
  \href{http://arxiv.org/abs/2003.10095}{{\ttfamily arXiv:2003.10095
  [astro-ph.CO]}}.

\bibitem{Chudaykin:2020acu}
A.~Chudaykin, D.~Gorbunov, and N.~Nedelko, ``{Combined analysis of Planck and
  SPTPol data favors the early dark energy models},''
  \href{http://arxiv.org/abs/2004.13046}{{\ttfamily arXiv:2004.13046
  [astro-ph.CO]}}.

\bibitem{Alestas:2020mvb}
G.~Alestas, L.~Kazantzidis, and L.~Perivolaropoulos, ``{$H_0$ Tension, Phantom
  Dark Energy and Cosmological Parameter Degeneracies},''
  \href{http://dx.doi.org/10.1103/PhysRevD.101.123516}{{\em Phys. Rev. D}
  {\bfseries 101} no.~12, (2020) 123516},
  \href{http://arxiv.org/abs/2004.08363}{{\ttfamily arXiv:2004.08363
  [astro-ph.CO]}}.

\bibitem{Clark:2020miy}
S.~J. Clark, K.~Vattis, and S.~M. Koushiappas, ``{CMB constraints on
  late-universe decaying dark matter as a solution to the $H_0$ tension},''
  \href{http://arxiv.org/abs/2006.03678}{{\ttfamily arXiv:2006.03678
  [astro-ph.CO]}}.

\bibitem{Keeley:2020rmo}
R.~E. Keeley, A.~Shafieloo, D.~K. Hazra, and T.~Souradeep, ``{Inflation Wars: A
  New Hope},'' \href{http://arxiv.org/abs/2006.12710}{{\ttfamily
  arXiv:2006.12710 [astro-ph.CO]}}.

\bibitem{Niedermann:2020dwg}
F.~Niedermann and M.~S. Sloth, ``{Resolving the Hubble Tension with New Early
  Dark Energy},'' \href{http://arxiv.org/abs/2006.06686}{{\ttfamily
  arXiv:2006.06686 [astro-ph.CO]}}.

\bibitem{Archidiacono:2020yey}
M.~Archidiacono, S.~Gariazzo, C.~Giunti, S.~Hannestad, and T.~Tram, ``{Sterile
  neutrino self-interactions: $H_0$ tension and short-baseline anomalies},''
  \href{http://arxiv.org/abs/2006.12885}{{\ttfamily arXiv:2006.12885
  [astro-ph.CO]}}.

\bibitem{DiValentino:2020kha}
E.~Di~Valentino, E.~V. Linder, and A.~Melchiorri, ``{$H_0$ Ex Machina: Vacuum
  Metamorphosis and Beyond $H_0$},''
  \href{http://arxiv.org/abs/2006.16291}{{\ttfamily arXiv:2006.16291
  [astro-ph.CO]}}.

\bibitem{Capozziello:2020nyq}
S.~Capozziello, M.~Benetti, and A.~D. Spallicci, ``{Addressing the cosmological
  $H_0$ tension by the Heisenberg uncertainty},''
  \href{http://arxiv.org/abs/2007.00462}{{\ttfamily arXiv:2007.00462 [gr-qc]}}.

\bibitem{Anchordoqui:2019yzc}
L.~A. Anchordoqui and S.~E. Perez~Bergliaffa, ``{Hot thermal universe endowed
  with massive dark vector fields and the Hubble tension},''
  \href{http://dx.doi.org/10.1103/PhysRevD.100.123525}{{\em Phys. Rev. D}
  {\bfseries 100} no.~12, (2019) 123525},
  \href{http://arxiv.org/abs/1910.05860}{{\ttfamily arXiv:1910.05860
  [astro-ph.CO]}}.

\bibitem{Ivanov:2020mfr}
M.~M. Ivanov, Y.~Ali-Haïmoud, and J.~Lesgourgues, ``{H0 tension or T0
  tension?},'' \href{http://arxiv.org/abs/2005.10656}{{\ttfamily
  arXiv:2005.10656 [astro-ph.CO]}}.

\bibitem{Gonzalez:2020fdy}
M.~Gonzalez, M.~P. Hertzberg, and F.~Rompineve, ``{Ultralight Scalar Decay and
  the Hubble Tension},'' \href{http://arxiv.org/abs/2006.13959}{{\ttfamily
  arXiv:2006.13959 [astro-ph.CO]}}.

\bibitem{Hryczuk:2020jhi}
A.~Hryczuk and K.~Jod\l~owski, ``{Self-interacting dark matter from late decays
  and the $H_0$ tension},'' \href{http://arxiv.org/abs/2006.16139}{{\ttfamily
  arXiv:2006.16139 [hep-ph]}}.

\bibitem{Baxter:2020qlr}
E.~J. Baxter and B.~D. Sherwin, ``{Determining the Hubble Constant without the
  Sound Horizon Scale: Measurements from CMB Lensing},''
  \href{http://arxiv.org/abs/2007.04007}{{\ttfamily arXiv:2007.04007
  [astro-ph.CO]}}.

\bibitem{Anchordoqui:2011nh}
L.~A. Anchordoqui and H.~Goldberg, ``{Neutrino cosmology after WMAP 7-Year data
  and LHC first Z' bounds},''
  \href{http://dx.doi.org/10.1103/PhysRevLett.108.081805}{{\em Phys. Rev.
  Lett.} {\bfseries 108} (2012) 081805},
  \href{http://arxiv.org/abs/1111.7264}{{\ttfamily arXiv:1111.7264 [hep-ph]}}.

\bibitem{Jacques:2013xr}
T.~D. Jacques, L.~M. Krauss, and C.~Lunardini, ``{Additional Light Sterile
  Neutrinos and Cosmology},''
  \href{http://dx.doi.org/10.1103/PhysRevD.87.083515}{{\em Phys. Rev. D}
  {\bfseries 87} no.~8, (2013) 083515},
  \href{http://arxiv.org/abs/1301.3119}{{\ttfamily arXiv:1301.3119
  [astro-ph.CO]}}. [Erratum: Phys.Rev.D 88, 109901 (2013)].

\bibitem{Weinberg:2013kea}
S.~Weinberg, ``{Goldstone Bosons as Fractional Cosmic Neutrinos},''
  \href{http://dx.doi.org/10.1103/PhysRevLett.110.241301}{{\em Phys. Rev.
  Lett.} {\bfseries 110} no.~24, (2013) 241301},
  \href{http://arxiv.org/abs/1305.1971}{{\ttfamily arXiv:1305.1971
  [astro-ph.CO]}}.

\bibitem{Anchordoqui:2012qu}
L.~A. Anchordoqui, H.~Goldberg, and G.~Steigman, ``{Right-Handed Neutrinos as
  the Dark Radiation: Status and Forecasts for the LHC},''
  \href{http://dx.doi.org/10.1016/j.physletb.2012.12.019}{{\em Phys. Lett. B}
  {\bfseries 718} (2013) 1162--1165},
  \href{http://arxiv.org/abs/1211.0186}{{\ttfamily arXiv:1211.0186 [hep-ph]}}.

\bibitem{Carneiro:2018xwq}
S.~Carneiro, P.~C. de~Holanda, C.~Pigozzo, and F.~Sobreira, ``{Is the $H_0$
  tension suggesting a fourth neutrino generation?},''
  \href{http://dx.doi.org/10.1103/PhysRevD.100.023505}{{\em Phys. Rev.}
  {\bfseries D100} no.~2, (2019) 023505},
\href{http://arxiv.org/abs/1812.06064}{{\ttfamily arXiv:1812.06064
  [astro-ph.CO]}}.

\bibitem{Paul:2018njm}
A.~Paul, A.~Ghoshal, A.~Chatterjee, and S.~Pal, ``{Inflation, (P)reheating and
  Neutrino Anomalies: Production of Sterile Neutrinos with Secret
  Interactions},'' \href{http://dx.doi.org/10.1140/epjc/s10052-019-7348-5}{{\em
  Eur. Phys. J.} {\bfseries C79} no.~10, (2019) 818},
\href{http://arxiv.org/abs/1808.09706}{{\ttfamily arXiv:1808.09706
  [astro-ph.CO]}}.

\bibitem{DiValentino:2015sam}
E.~Di~Valentino, E.~Giusarma, O.~Mena, A.~Melchiorri, and J.~Silk,
  ``{Cosmological limits on neutrino unknowns versus low redshift priors},''
  \href{http://dx.doi.org/10.1103/PhysRevD.93.083527}{{\em Phys. Rev.}
  {\bfseries D93} no.~8, (2016) 083527},
\href{http://arxiv.org/abs/1511.00975}{{\ttfamily arXiv:1511.00975
  [astro-ph.CO]}}.

\bibitem{Green:2019glg}
D.~Green {\em et~al.}, ``{Messengers from the Early Universe: Cosmic Neutrinos
  and Other Light Relics},'' {\em Bull. Am. Astron. Soc.} {\bfseries 51} no.~7,
  (2019) 159, \href{http://arxiv.org/abs/1903.04763}{{\ttfamily
  arXiv:1903.04763 [astro-ph.CO]}}.

\bibitem{Barenboim:2016lxv}
G.~Barenboim, W.~H. Kinney, and W.-I. Park, ``{Flavor versus mass eigenstates
  in neutrino asymmetries: implications for cosmology},''
  \href{http://dx.doi.org/10.1140/epjc/s10052-017-5147-4}{{\em Eur. Phys. J. C}
  {\bfseries 77} no.~9, (2017) 590},
  \href{http://arxiv.org/abs/1609.03200}{{\ttfamily arXiv:1609.03200
  [astro-ph.CO]}}.

\bibitem{Ferreira:2018vjj}
R.~Z. Ferreira and A.~Notari, ``{Observable Windows for the QCD Axion Through
  the Number of Relativistic Species},''
  \href{http://dx.doi.org/10.1103/PhysRevLett.120.191301}{{\em Phys. Rev.
  Lett.} {\bfseries 120} no.~19, (2018) 191301},
\href{http://arxiv.org/abs/1801.06090}{{\ttfamily arXiv:1801.06090 [hep-ph]}}.

\bibitem{Gelmini:2019deq}
G.~B. Gelmini, A.~Kusenko, and V.~Takhistov, ``{Hints of Sterile Neutrinos in
  Recent Measurements of the Hubble Parameter},''
  \href{http://arxiv.org/abs/1906.10136}{{\ttfamily arXiv:1906.10136
  [astro-ph.CO]}}.

\bibitem{DiValentino:2015wba}
E.~Di~Valentino, E.~Giusarma, M.~Lattanzi, O.~Mena, A.~Melchiorri, and J.~Silk,
  ``{Cosmological Axion and neutrino mass constraints from Planck 2015
  temperature and polarization data},''
  \href{http://dx.doi.org/10.1016/j.physletb.2015.11.025}{{\em Phys. Lett.}
  {\bfseries B752} (2016) 182--185},
\href{http://arxiv.org/abs/1507.08665}{{\ttfamily arXiv:1507.08665
  [astro-ph.CO]}}.

\bibitem{Poulin:2018dzj}
V.~Poulin, T.~L. Smith, D.~Grin, T.~Karwal, and M.~Kamionkowski,
  ``{Cosmological implications of ultralight axionlike fields},''
  \href{http://dx.doi.org/10.1103/PhysRevD.98.083525}{{\em Phys. Rev. D}
  {\bfseries 98} no.~8, (2018) 083525},
  \href{http://arxiv.org/abs/1806.10608}{{\ttfamily arXiv:1806.10608
  [astro-ph.CO]}}.

\bibitem{Baumann:2016wac}
D.~Baumann, D.~Green, and B.~Wallisch, ``{New Target for Cosmic Axion
  Searches},'' \href{http://dx.doi.org/10.1103/PhysRevLett.117.171301}{{\em
  Phys. Rev. Lett.} {\bfseries 117} no.~17, (2016) 171301},
  \href{http://arxiv.org/abs/1604.08614}{{\ttfamily arXiv:1604.08614
  [astro-ph.CO]}}.

\bibitem{Anchordoqui:2020znj}
L.~A. Anchordoqui, ``{Hubble Hullabaloo and String Cosmology},''
\newblock 5, 2020.
\newblock \href{http://arxiv.org/abs/2005.01217}{{\ttfamily arXiv:2005.01217
  [astro-ph.CO]}}.

\bibitem{Sakstein:2019fmf}
J.~Sakstein and M.~Trodden, ``{Early Dark Energy from Massive Neutrinos as a
  Natural Resolution of the Hubble Tension},''
  \href{http://dx.doi.org/10.1103/PhysRevLett.124.161301}{{\em Phys. Rev.
  Lett.} {\bfseries 124} no.~16, (2020) 161301},
  \href{http://arxiv.org/abs/1911.11760}{{\ttfamily arXiv:1911.11760
  [astro-ph.CO]}}.

\bibitem{Das:2020wfe}
A.~Gogoi, P.~Chanda, and S.~Das, ``{Dark matter nugget and new early dark
  energy from interacting neutrino: A promising solution to Hubble anomaly},''
  \href{http://arxiv.org/abs/2005.11889}{{\ttfamily arXiv:2005.11889
  [astro-ph.CO]}}.

\bibitem{Akarsu:2019hmw}
{\"O}.~Akarsu, J.~D. Barrow, L.~A. Escamilla, and J.~A. Vazquez, ``{Graduated
  dark energy: Observational hints of a spontaneous sign switch in the
  cosmological constant},''
  \href{http://dx.doi.org/10.1103/PhysRevD.101.063528}{{\em Phys. Rev. D}
  {\bfseries 101} no.~6, (2020) 063528},
  \href{http://arxiv.org/abs/1912.08751}{{\ttfamily arXiv:1912.08751
  [astro-ph.CO]}}.

\bibitem{Ye:2020btb}
G.~Ye and Y.-S. Piao, ``{Is the Hubble tension a hint of AdS around
  recombination?},''
\href{http://arxiv.org/abs/2001.02451}{{\ttfamily arXiv:2001.02451
  [astro-ph.CO]}}.

\bibitem{Hart:2017ndk}
L.~Hart and J.~Chluba, ``{New constraints on time-dependent variations of
  fundamental constants using Planck data},''
  \href{http://dx.doi.org/10.1093/mnras/stx2783}{{\em Mon. Not. Roy. Astron.
  Soc.} {\bfseries 474} no.~2, (2018) 1850--1861},
  \href{http://arxiv.org/abs/1705.03925}{{\ttfamily arXiv:1705.03925
  [astro-ph.CO]}}.

\bibitem{Chiang:2018xpn}
C.-T. Chiang and A.~z. Slosar, ``{Inferences of $H_0$ in presence of a
  non-standard recombination},''
  \href{http://arxiv.org/abs/1811.03624}{{\ttfamily arXiv:1811.03624
  [astro-ph.CO]}}.

\bibitem{Hart:2019dxi}
L.~Hart and J.~Chluba, ``{Updated fundamental constant constraints from Planck
  2018 data and possible relations to the Hubble tension},''
  \href{http://dx.doi.org/10.1093/mnras/staa412}{{\em Mon. Not. Roy. Astron.
  Soc.} {\bfseries 493} no.~3, (2020) 3255--3263},
  \href{http://arxiv.org/abs/1912.03986}{{\ttfamily arXiv:1912.03986
  [astro-ph.CO]}}.

\bibitem{Jedamzik:2020krr}
K.~Jedamzik and L.~Pogosian, ``{Relieving the Hubble tension with primordial
  magnetic fields},'' \href{http://arxiv.org/abs/2004.09487}{{\ttfamily
  arXiv:2004.09487 [astro-ph.CO]}}.

\bibitem{Sekiguchi:2020teg}
T.~Sekiguchi and T.~Takahashi, ``{Early recombination as a solution to the
  $H_0$ tension},'' \href{http://arxiv.org/abs/2007.03381}{{\ttfamily
  arXiv:2007.03381 [astro-ph.CO]}}.

\bibitem{Bose:2020cjb}
B.~Bose and L.~Lombriser, ``{Easing cosmic tensions with an open and hotter
  universe},'' \href{http://arxiv.org/abs/2006.16149}{{\ttfamily
  arXiv:2006.16149 [astro-ph.CO]}}.

\bibitem{Agrawal:2019dlm}
P.~Agrawal, G.~Obied, and C.~Vafa, ``{$H_0$ Tension, Swampland Conjectures and
  the Epoch of Fading Dark Matter},''
  \href{http://arxiv.org/abs/1906.08261}{{\ttfamily arXiv:1906.08261
  [astro-ph.CO]}}.

\bibitem{Anchordoqui:2019amx}
L.~A. Anchordoqui, I.~Antoniadis, D.~Lüst, J.~F. Soriano, and T.~R. Taylor,
  ``{$H_0$ tension and the String Swampland},''
  \href{http://dx.doi.org/10.1103/PhysRevD.101.083532}{{\em Phys. Rev. D}
  {\bfseries 101} (2020) 083532},
  \href{http://arxiv.org/abs/1912.00242}{{\ttfamily arXiv:1912.00242
  [hep-th]}}.

\bibitem{Anchordoqui:2020sqo}
L.~A. Anchordoqui, I.~Antoniadis, D.~Lüst, and J.~F. Soriano, ``{Dark energy,
  Ricci-nonflat spaces, and the Swampland},''
  \href{http://arxiv.org/abs/2005.10075}{{\ttfamily arXiv:2005.10075
  [hep-th]}}.

\bibitem{Ade:2015rim}
{\bfseries Planck} Collaboration, P.~A.~R. Ade {\em et~al.}, ``{Planck 2015
  results. XIV. Dark energy and modified gravity},''
  \href{http://dx.doi.org/10.1051/0004-6361/201525814}{{\em Astron. Astrophys.}
  {\bfseries 594} (2016) A14},
\href{http://arxiv.org/abs/1502.01590}{{\ttfamily arXiv:1502.01590
  [astro-ph.CO]}}.

\bibitem{Raveri:2019mxg}
M.~Raveri, ``{Reconstructing Gravity on Cosmological Scales},''
  \href{http://dx.doi.org/10.1103/PhysRevD.101.083524}{{\em Phys. Rev. D}
  {\bfseries 101} no.~8, (2020) 083524},
  \href{http://arxiv.org/abs/1902.01366}{{\ttfamily arXiv:1902.01366
  [astro-ph.CO]}}.

\bibitem{Yan:2019gbw}
S.-F. Yan, P.~Zhang, J.-W. Chen, X.-Z. Zhang, Y.-F. Cai, and E.~N. Saridakis,
  ``{Interpreting cosmological tensions from the effective field theory of
  torsional gravity},''
  \href{http://dx.doi.org/10.1103/PhysRevD.101.121301}{{\em Phys. Rev. D}
  {\bfseries 101} no.~12, (2020) 121301},
  \href{http://arxiv.org/abs/1909.06388}{{\ttfamily arXiv:1909.06388
  [astro-ph.CO]}}.

\bibitem{Frusciante:2019puu}
N.~Frusciante, S.~Peirone, L.~Atayde, and A.~De~Felice, ``{Phenomenology of the
  generalized cubic covariant Galileon model and cosmological bounds},''
  \href{http://dx.doi.org/10.1103/PhysRevD.101.064001}{{\em Phys. Rev. D}
  {\bfseries 101} no.~6, (2020) 064001},
  \href{http://arxiv.org/abs/1912.07586}{{\ttfamily arXiv:1912.07586
  [astro-ph.CO]}}.

\bibitem{Braglia:2020iik}
M.~Braglia, M.~Ballardini, W.~T. Emond, F.~Finelli, A.~E. Gumrukcuoglu,
  K.~Koyama, and D.~Paoletti, ``{A larger value for $H_0$ by an evolving
  gravitational constant},'' \href{http://arxiv.org/abs/2004.11161}{{\ttfamily
  arXiv:2004.11161 [astro-ph.CO]}}.

\bibitem{Ballardini:2020iws}
M.~Ballardini, M.~Braglia, F.~Finelli, D.~Paoletti, A.~A. Starobinsky, and
  C.~Umiltà, ``{Scalar-tensor theories of gravity, neutrino physics, and the
  $H_0$ tension},'' \href{http://arxiv.org/abs/2004.14349}{{\ttfamily
  arXiv:2004.14349 [astro-ph.CO]}}.

\bibitem{Rossi:2019lgt}
M.~Rossi, M.~Ballardini, M.~Braglia, F.~Finelli, D.~Paoletti, A.~A.
  Starobinsky, and C.~Umiltà, ``{Cosmological constraints on post-Newtonian
  parameters in effectively massless scalar-tensor theories of gravity},''
  \href{http://dx.doi.org/10.1103/PhysRevD.100.103524}{{\em Phys. Rev. D}
  {\bfseries 100} no.~10, (2019) 103524},
  \href{http://arxiv.org/abs/1906.10218}{{\ttfamily arXiv:1906.10218
  [astro-ph.CO]}}.

\bibitem{Pan:2019hac}
S.~Pan, W.~Yang, E.~Di~Valentino, A.~Shafieloo, and S.~Chakraborty,
  ``{Reconciling $H_0$ tension in a six parameter space?},''
  \href{http://dx.doi.org/10.1088/1475-7516/2020/06/062}{{\em JCAP} {\bfseries
  06} (2020) 062}, \href{http://arxiv.org/abs/1907.12551}{{\ttfamily
  arXiv:1907.12551 [astro-ph.CO]}}.

\bibitem{Li:2019yem}
X.~Li and A.~Shafieloo, ``{A Simple Phenomenological Emergent Dark Energy Model
  can Resolve the Hubble Tension},''
  \href{http://dx.doi.org/10.3847/2041-8213/ab3e09}{{\em Astrophys. J. Lett.}
  {\bfseries 883} no.~1, (2019) L3},
  \href{http://arxiv.org/abs/1906.08275}{{\ttfamily arXiv:1906.08275
  [astro-ph.CO]}}.

\bibitem{Rezaei:2020mrj}
M.~Rezaei, T.~Naderi, M.~Malekjani, and A.~Mehrabi, ``{A Bayesian comparison
  between $\Lambda$CDM and phenomenologically emergent dark energy models},''
  \href{http://dx.doi.org/10.1140/epjc/s10052-020-7942-6}{{\em Eur. Phys. J. C}
  {\bfseries 80} no.~5, (2020) 374},
  \href{http://arxiv.org/abs/2004.08168}{{\ttfamily arXiv:2004.08168
  [astro-ph.CO]}}.

\bibitem{Liu:2020vgn}
Z.~Liu and H.~Miao, ``{Neutrino mass and mass hierarchy in various dark
  energy},'' \href{http://arxiv.org/abs/2002.05563}{{\ttfamily arXiv:2002.05563
  [astro-ph.CO]}}.

\bibitem{Li:2020ybr}
X.~Li and A.~Shafieloo, ``{Generalised Emergent Dark Energy Model: Confronting
  $\Lambda$ and PEDE},'' \href{http://arxiv.org/abs/2001.05103}{{\ttfamily
  arXiv:2001.05103 [astro-ph.CO]}}.

\bibitem{Yang:2020tax}
W.~Yang, E.~Di~Valentino, O.~Mena, and S.~Pan, ``{Dynamical Dark sectors and
  Neutrino masses and abundances},''
  \href{http://dx.doi.org/10.1103/PhysRevD.102.023535}{{\em Phys. Rev. D}
  {\bfseries 102} no.~2, (2020) 023535},
  \href{http://arxiv.org/abs/2003.12552}{{\ttfamily arXiv:2003.12552
  [astro-ph.CO]}}.

\bibitem{DiBari:2013dna}
P.~Di~Bari, S.~F. King, and A.~Merle, ``{Dark Radiation or Warm Dark Matter
  from long lived particle decays in the light of Planck},''
  \href{http://dx.doi.org/10.1016/j.physletb.2013.06.003}{{\em Phys. Lett. B}
  {\bfseries 724} (2013) 77--83},
  \href{http://arxiv.org/abs/1303.6267}{{\ttfamily arXiv:1303.6267 [hep-ph]}}.

\bibitem{Choi:2019jck}
G.~Choi, M.~Suzuki, and T.~T. Yanagida, ``{Quintessence Axion Dark Energy and a
  Solution to the Hubble Tension},''
  \href{http://dx.doi.org/10.1016/j.physletb.2020.135408}{{\em Phys. Lett. B}
  {\bfseries 805} (2020) 135408},
  \href{http://arxiv.org/abs/1910.00459}{{\ttfamily arXiv:1910.00459
  [hep-ph]}}.

\bibitem{Choi:2020tqp}
G.~Choi, M.~Suzuki, and T.~T. Yanagida, ``{Degenerate Sub-keV Fermion Dark
  Matter from a Solution to the Hubble Tension},''
  \href{http://dx.doi.org/10.1103/PhysRevD.101.075031}{{\em Phys. Rev. D}
  {\bfseries 101} no.~7, (2020) 075031},
  \href{http://arxiv.org/abs/2002.00036}{{\ttfamily arXiv:2002.00036
  [hep-ph]}}.

\bibitem{Berezhiani:2015yta}
Z.~Berezhiani, A.~Dolgov, and I.~Tkachev, ``{Reconciling Planck results with
  low redshift astronomical measurements},''
  \href{http://dx.doi.org/10.1103/PhysRevD.92.061303}{{\em Phys. Rev. D}
  {\bfseries 92} no.~6, (2015) 061303},
  \href{http://arxiv.org/abs/1505.03644}{{\ttfamily arXiv:1505.03644
  [astro-ph.CO]}}.

\bibitem{Anchordoqui:2015lqa}
L.~A. Anchordoqui, V.~Barger, H.~Goldberg, X.~Huang, D.~Marfatia, L.~H.~M.
  da~Silva, and T.~J. Weiler, ``{IceCube neutrinos, decaying dark matter, and
  the Hubble constant},''
  \href{http://dx.doi.org/10.1103/PhysRevD.94.069901}{{\em Phys. Rev. D}
  {\bfseries 92} no.~6, (2015) 061301},
  \href{http://arxiv.org/abs/1506.08788}{{\ttfamily arXiv:1506.08788
  [hep-ph]}}. [Erratum: Phys.Rev.D 94, 069901 (2016)].

\bibitem{Vattis:2019efj}
K.~Vattis, S.~M. Koushiappas, and A.~Loeb, ``{Dark matter decaying in the late
  Universe can relieve the H0 tension},''
  \href{http://dx.doi.org/10.1103/PhysRevD.99.121302}{{\em Phys. Rev.}
  {\bfseries D99} no.~12, (2019) 121302},
\href{http://arxiv.org/abs/1903.06220}{{\ttfamily arXiv:1903.06220
  [astro-ph.CO]}}.

\bibitem{Desai:2019pvs}
A.~Desai, K.~R. Dienes, and B.~Thomas, ``{Constraining Dark-Matter Ensembles
  with Supernova Data},''
  \href{http://dx.doi.org/10.1103/PhysRevD.101.035031}{{\em Phys. Rev. D}
  {\bfseries 101} no.~3, (2020) 035031},
  \href{http://arxiv.org/abs/1909.07981}{{\ttfamily arXiv:1909.07981
  [astro-ph.CO]}}.

\bibitem{Alcaniz:2019kah}
J.~Alcaniz, N.~Bernal, A.~Masiero, and F.~S. Queiroz, ``{Light Dark Matter: A
  Common Solution to the Lithium and ${H_0}$ Problems},''
  \href{http://arxiv.org/abs/1912.05563}{{\ttfamily arXiv:1912.05563
  [astro-ph.CO]}}.

\bibitem{Chudaykin:2016yfk}
A.~Chudaykin, D.~Gorbunov, and I.~Tkachev, ``{Dark matter component decaying
  after recombination: Lensing constraints with Planck data},''
  \href{http://dx.doi.org/10.1103/PhysRevD.94.023528}{{\em Phys. Rev. D}
  {\bfseries 94} (2016) 023528},
  \href{http://arxiv.org/abs/1602.08121}{{\ttfamily arXiv:1602.08121
  [astro-ph.CO]}}.

\bibitem{Chudaykin:2017ptd}
A.~Chudaykin, D.~Gorbunov, and I.~Tkachev, ``{Dark matter component decaying
  after recombination: Sensitivity to baryon acoustic oscillation and redshift
  space distortion probes},''
  \href{http://dx.doi.org/10.1103/PhysRevD.97.083508}{{\em Phys. Rev. D}
  {\bfseries 97} no.~8, (2018) 083508},
  \href{http://arxiv.org/abs/1711.06738}{{\ttfamily arXiv:1711.06738
  [astro-ph.CO]}}.

\bibitem{Hill:2020osr}
J.~C. Hill, E.~McDonough, M.~W. Toomey, and S.~Alexander, ``{Early Dark Energy
  Does Not Restore Cosmological Concordance},''
  \href{http://arxiv.org/abs/2003.07355}{{\ttfamily arXiv:2003.07355
  [astro-ph.CO]}}.

\bibitem{Ivanov:2020ril}
M.~M. Ivanov, E.~McDonough, J.~C. Hill, M.~Simonovi\'c, M.~W. Toomey,
  S.~Alexander, and M.~Zaldarriaga, ``{Constraining Early Dark Energy with
  Large-Scale Structure},'' \href{http://arxiv.org/abs/2006.11235}{{\ttfamily
  arXiv:2006.11235 [astro-ph.CO]}}.

\bibitem{Rezaei:2020lfy}
M.~Rezaei, S.~P. Ojaghi, and M.~Malekjani, ``{Cosmography approach to dark
  energy cosmologies: new constrains using the Hubble diagrams of supernovae,
  quasars and gamma-ray bursts},''
  \href{http://arxiv.org/abs/2008.03092}{{\ttfamily arXiv:2008.03092
  [astro-ph.CO]}}.

\bibitem{Wang:2020dsc}
D.~Wang, ``{Can $f(R)$ gravity relieve $H_0$ and $\sigma_8$ tensions?},''
  \href{http://arxiv.org/abs/2008.03966}{{\ttfamily arXiv:2008.03966
  [astro-ph.CO]}}.

\bibitem{Nunes:2020uex}
R.~C. Nunes and A.~Bernui, ``{$\theta_{\rm BAO}$ estimates and the $H_0$
  tension},'' \href{http://arxiv.org/abs/2008.03259}{{\ttfamily
  arXiv:2008.03259 [astro-ph.CO]}}.

\bibitem{Leonhardt:2020qam}
U.~Leonhardt and D.~Berechya, ``{Observed Hubble constant is consistent with
  physics of the quantum vacuum},''
  \href{http://arxiv.org/abs/2008.04789}{{\ttfamily arXiv:2008.04789 [gr-qc]}}.

\bibitem{Birrer:2020jyr}
S.~Birrer and T.~Treu, ``{TDCOSMO V: strategies for precise and accurate
  measurements of the Hubble constant with strong lensing},''
  \href{http://arxiv.org/abs/2008.06157}{{\ttfamily arXiv:2008.06157
  [astro-ph.CO]}}.

\bibitem{Ballesteros:2020sik}
G.~Ballesteros, A.~Notari, and F.~Rompineve, ``{The $H_0$ tension: $\Delta G_N$
  vs. $\Delta N_{\rm eff}$},''
  \href{http://arxiv.org/abs/2004.05049}{{\ttfamily arXiv:2004.05049
  [astro-ph.CO]}}.

\bibitem{Blinov:2019gcj}
N.~Blinov, K.~J. Kelly, G.~Z. Krnjaic, and S.~D. McDermott, ``{Constraining the
  Self-Interacting Neutrino Interpretation of the Hubble Tension},''
  \href{http://dx.doi.org/10.1103/PhysRevLett.123.191102}{{\em Phys. Rev.
  Lett.} {\bfseries 123} no.~19, (2019) 191102},
  \href{http://arxiv.org/abs/1905.02727}{{\ttfamily arXiv:1905.02727
  [astro-ph.CO]}}.

\bibitem{Hernandez-Almada:2020uyr}
A.~Hernández-Almada, G.~Leon, J.~Magaña, M.~A. García-Aspeitia, and
  V.~Motta, ``{Generalized Emergent Dark Energy: observational Hubble data
  constraints and stability analysis},''
  \href{http://arxiv.org/abs/2002.12881}{{\ttfamily arXiv:2002.12881
  [astro-ph.CO]}}.

\bibitem{Philcox:2020xbv}
O.~H. Philcox, B.~D. Sherwin, G.~S. Farren, and E.~J. Baxter, ``{Determining
  the Hubble Constant without the Sound Horizon: Measurements from Galaxy
  Surveys},'' \href{http://arxiv.org/abs/2008.08084}{{\ttfamily
  arXiv:2008.08084 [astro-ph.CO]}}.

\bibitem{Feeney:2017sgx}
S.~M. Feeney, D.~J. Mortlock, and N.~Dalmasso, ``{Clarifying the Hubble
  constant tension with a Bayesian hierarchical model of the local distance
  ladder},'' \href{http://dx.doi.org/10.1093/mnras/sty418}{{\em Mon. Not. Roy.
  Astron. Soc.} {\bfseries 476} no.~3, (2018) 3861--3882},
  \href{http://arxiv.org/abs/1707.00007}{{\ttfamily arXiv:1707.00007
  [astro-ph.CO]}}.

\bibitem{Feeney:2018mkj}
S.~M. Feeney, H.~V. Peiris, A.~R. Williamson, S.~M. Nissanke, D.~J. Mortlock,
  J.~Alsing, and D.~Scolnic, ``{Prospects for resolving the Hubble constant
  tension with standard sirens},''
  \href{http://dx.doi.org/10.1103/PhysRevLett.122.061105}{{\em Phys. Rev.
  Lett.} {\bfseries 122} no.~6, (2019) 061105},
  \href{http://arxiv.org/abs/1802.03404}{{\ttfamily arXiv:1802.03404
  [astro-ph.CO]}}.

\bibitem{Mortlock:2018azx}
D.~J. Mortlock, S.~M. Feeney, H.~V. Peiris, A.~R. Williamson, and S.~M.
  Nissanke, ``{Unbiased Hubble constant estimation from binary neutron star
  mergers},'' \href{http://dx.doi.org/10.1103/PhysRevD.100.103523}{{\em Phys.
  Rev. D} {\bfseries 100} no.~10, (2019) 103523},
  \href{http://arxiv.org/abs/1811.11723}{{\ttfamily arXiv:1811.11723
  [astro-ph.CO]}}.

\bibitem{Banerjee:2020xcn}
A.~Banerjee, H.~Cai, L.~Heisenberg, E.~O. Colg\'ain, M.~Sheikh-Jabbari, and
  T.~Yang, ``{Hubble Sinks In The Low-Redshift Swampland},''
  \href{http://arxiv.org/abs/2006.00244}{{\ttfamily arXiv:2006.00244
  [astro-ph.CO]}}.

\bibitem{Adler:2019fnp}
S.~L. Adler, ``{Implications of a frame dependent dark energy for the spacetime
  metric, cosmography, and effective Hubble constant},''
  \href{http://dx.doi.org/10.1103/PhysRevD.100.123503}{{\em Phys. Rev. D}
  {\bfseries 100} no.~12, (2019) 123503},
  \href{http://arxiv.org/abs/1905.08228}{{\ttfamily arXiv:1905.08228
  [astro-ph.CO]}}.

\bibitem{Gu:2020ozv}
Y.~Gu, M.~Khlopov, L.~Wu, J.~M. Yang, and B.~Zhu, ``{Light gravitino dark
  matter for Hubble tension and LHC},''
  \href{http://arxiv.org/abs/2006.09906}{{\ttfamily arXiv:2006.09906
  [hep-ph]}}.

\bibitem{Akarsu:2019pwn}
{\"O}.~Akarsu, S.~Kumar, S.~Sharma, and L.~Tedesco, ``{Constraints on a Bianchi
  type I spacetime extension of the standard $\Lambda$CDM model},''
  \href{http://dx.doi.org/10.1103/PhysRevD.100.023532}{{\em Phys. Rev. D}
  {\bfseries 100} no.~2, (2019) 023532},
  \href{http://arxiv.org/abs/1905.06949}{{\ttfamily arXiv:1905.06949
  [astro-ph.CO]}}.

\bibitem{DiValentino:2020vhf}
E.~Di~Valentino {\em et~al.}, ``{Cosmology Intertwined I: Perspectives for the
  Next Decade},'' \href{http://arxiv.org/abs/2008.11283}{{\ttfamily
  arXiv:2008.11283 [astro-ph.CO]}}.

\bibitem{DiValentino:2020vvd}
E.~Di~Valentino {\em et~al.}, ``{Cosmology Intertwined III: $f \sigma_8$ and
  $S_8$},'' \href{http://arxiv.org/abs/2008.11285}{{\ttfamily arXiv:2008.11285
  [astro-ph.CO]}}.

\bibitem{DiValentino:2020srs}
E.~Di~Valentino {\em et~al.}, ``{Cosmology Intertwined IV: The Age of the
  Universe and its Curvature},''
  \href{http://arxiv.org/abs/2008.11286}{{\ttfamily arXiv:2008.11286
  [astro-ph.CO]}}.

\bibitem{Pettorino:2013ia}
V.~Pettorino, L.~Amendola, and C.~Wetterich, ``{How early is early dark
  energy?},'' \href{http://dx.doi.org/10.1103/PhysRevD.87.083009}{{\em Phys.
  Rev. D} {\bfseries 87} (2013) 083009},
  \href{http://arxiv.org/abs/1301.5279}{{\ttfamily arXiv:1301.5279
  [astro-ph.CO]}}.

\bibitem{Pettorino:2013oxa}
V.~Pettorino, ``{Testing modified gravity with Planck: the case of coupled dark
  energy},'' \href{http://dx.doi.org/10.1103/PhysRevD.88.063519}{{\em Phys.
  Rev. D} {\bfseries 88} (2013) 063519},
  \href{http://arxiv.org/abs/1305.7457}{{\ttfamily arXiv:1305.7457
  [astro-ph.CO]}}.

\bibitem{Yang:2020uga}
W.~Yang, E.~Di~Valentino, O.~Mena, S.~Pan, and R.~C. Nunes, ``{All-inclusive
  interacting dark sector cosmologies},''
  \href{http://dx.doi.org/10.1103/PhysRevD.101.083509}{{\em Phys. Rev. D}
  {\bfseries 101} no.~8, (2020) 083509},
  \href{http://arxiv.org/abs/2001.10852}{{\ttfamily arXiv:2001.10852
  [astro-ph.CO]}}.

\bibitem{Yang:2019uog}
W.~Yang, S.~Pan, R.~C. Nunes, and D.~F. Mota, ``{Dark calling Dark: Interaction
  in the dark sector in presence of neutrino properties after Planck CMB final
  release},'' \href{http://dx.doi.org/10.1088/1475-7516/2020/04/008}{{\em JCAP}
  {\bfseries 04} (2020) 008}, \href{http://arxiv.org/abs/1910.08821}{{\ttfamily
  arXiv:1910.08821 [astro-ph.CO]}}.

\bibitem{Yang:2018ubt}
W.~Yang, S.~Pan, L.~Xu, and D.~F. Mota, ``{Effects of anisotropic stress in
  interacting dark matter -- dark energy scenarios},''
  \href{http://dx.doi.org/10.1093/mnras/sty2789}{{\em Mon. Not. Roy. Astron.
  Soc.} {\bfseries 482} no.~2, (2019) 1858--1871},
  \href{http://arxiv.org/abs/1804.08455}{{\ttfamily arXiv:1804.08455
  [astro-ph.CO]}}.

\bibitem{Steigman:1977kc}
G.~Steigman, D.~Schramm, and J.~Gunn, ``{Cosmological Limits to the Number of
  Massive Leptons},''
  \href{http://dx.doi.org/10.1016/0370-2693(77)90176-9}{{\em Phys. Lett. B}
  {\bfseries 66} (1977) 202--204}.

\bibitem{Mangano:2005cc}
G.~Mangano, G.~Miele, S.~Pastor, T.~Pinto, O.~Pisanti, and P.~D. Serpico,
  ``{Relic neutrino decoupling including flavor oscillations},''
  \href{http://dx.doi.org/10.1016/j.nuclphysb.2005.09.041}{{\em Nucl. Phys.}
  {\bfseries B729} (2005) 221--234},
\href{http://arxiv.org/abs/hep-ph/0506164}{{\ttfamily arXiv:hep-ph/0506164
  [hep-ph]}}.

\bibitem{deSalas:2016ztq}
P.~F. de~Salas and S.~Pastor, ``{Relic neutrino decoupling with flavour
  oscillations revisited},''
  \href{http://dx.doi.org/10.1088/1475-7516/2016/07/051}{{\em JCAP} {\bfseries
  1607} no.~07, (2016) 051},
\href{http://arxiv.org/abs/1606.06986}{{\ttfamily arXiv:1606.06986 [hep-ph]}}.

\bibitem{Akita:2020szl}
K.~Akita and M.~Yamaguchi, ``{A precision calculation of relic neutrino
  decoupling},'' \href{http://dx.doi.org/10.1088/1475-7516/2020/08/012}{{\em
  JCAP} {\bfseries 08} (2020) 012},
  \href{http://arxiv.org/abs/2005.07047}{{\ttfamily arXiv:2005.07047
  [hep-ph]}}.

\bibitem{Zeng:2018pcv}
Z.~Zeng, S.~Yeung, and M.-C. Chu, ``{Effects of neutrino mass and asymmetry on
  cosmological structure formation},''
  \href{http://dx.doi.org/10.1088/1475-7516/2019/03/015}{{\em JCAP} {\bfseries
  03} (2019) 015}, \href{http://arxiv.org/abs/1808.00357}{{\ttfamily
  arXiv:1808.00357 [astro-ph.CO]}}.

\bibitem{Allahverdi:2014ppa}
R.~Allahverdi, M.~Cicoli, B.~Dutta, and K.~Sinha, ``{Correlation between Dark
  Matter and Dark Radiation in String Compactifications},''
  \href{http://dx.doi.org/10.1088/1475-7516/2014/10/002}{{\em JCAP} {\bfseries
  10} (2014) 002}, \href{http://arxiv.org/abs/1401.4364}{{\ttfamily
  arXiv:1401.4364 [hep-ph]}}.

\bibitem{Abazajian:2016yjj}
{\bfseries CMB-S4} Collaboration, K.~N. Abazajian {\em et~al.}, ``{CMB-S4
  Science Book, First Edition},''
  \href{http://arxiv.org/abs/1610.02743}{{\ttfamily arXiv:1610.02743
  [astro-ph.CO]}}.

\bibitem{Abazajian:2019eic}
K.~Abazajian {\em et~al.}, ``{CMB-S4 Science Case, Reference Design, and
  Project Plan},'' \href{http://arxiv.org/abs/1907.04473}{{\ttfamily
  arXiv:1907.04473 [astro-ph.IM]}}.

\bibitem{Sola:2019jek}
J.~Solà~Peracaula, A.~Gómez-Valent, J.~de~Cruz~Pérez, and C.~Moreno-Pulido,
  ``{Brans--Dicke Gravity with a Cosmological Constant Smoothes Out
  $\Lambda$CDM Tensions},''
  \href{http://dx.doi.org/10.3847/2041-8213/ab53e9}{{\em Astrophys. J. Lett.}
  {\bfseries 886} no.~1, (2019) L6},
  \href{http://arxiv.org/abs/1909.02554}{{\ttfamily arXiv:1909.02554
  [astro-ph.CO]}}.

\bibitem{Sola:2020lba}
J.~Solà, A.~Gómez-Valent, J.~de~Cruz~Pérez, and C.~Moreno-Pulido,
  ``{Brans-Dicke cosmology with a $\Lambda$- term: a possible solution to
  $\Lambda$CDM tensions},'' \href{http://arxiv.org/abs/2006.04273}{{\ttfamily
  arXiv:2006.04273 [astro-ph.CO]}}.

\bibitem{Umilta:2015cta}
C.~Umiltà, M.~Ballardini, F.~Finelli, and D.~Paoletti, ``{CMB and BAO
  constraints for an induced gravity dark energy model with a quartic
  potential},'' \href{http://dx.doi.org/10.1088/1475-7516/2015/08/017}{{\em
  JCAP} {\bfseries 08} (2015) 017},
  \href{http://arxiv.org/abs/1507.00718}{{\ttfamily arXiv:1507.00718
  [astro-ph.CO]}}.

\bibitem{Ballardini:2016cvy}
M.~Ballardini, F.~Finelli, C.~Umiltà, and D.~Paoletti, ``{Cosmological
  constraints on induced gravity dark energy models},''
  \href{http://dx.doi.org/10.1088/1475-7516/2016/05/067}{{\em JCAP} {\bfseries
  05} (2016) 067}, \href{http://arxiv.org/abs/1601.03387}{{\ttfamily
  arXiv:1601.03387 [astro-ph.CO]}}.

\bibitem{Sahni:2014ooa}
V.~Sahni, A.~Shafieloo, and A.~A. Starobinsky, ``{Model independent evidence
  for dark energy evolution from Baryon Acoustic Oscillations},''
  \href{http://dx.doi.org/10.1088/2041-8205/793/2/L40}{{\em Astrophys. J.
  Lett.} {\bfseries 793} no.~2, (2014) L40},
  \href{http://arxiv.org/abs/1406.2209}{{\ttfamily arXiv:1406.2209
  [astro-ph.CO]}}.

\bibitem{Capozziello:2019cav}
S.~Capozziello, R.~D'Agostino, and O.~Luongo, ``{Extended Gravity
  Cosmography},'' \href{http://dx.doi.org/10.1142/S0218271819300167}{{\em Int.
  J. Mod. Phys. D} {\bfseries 28} no.~10, (2019) 1930016},
  \href{http://arxiv.org/abs/1904.01427}{{\ttfamily arXiv:1904.01427 [gr-qc]}}.

\bibitem{Benetti:2019gmo}
M.~Benetti and S.~Capozziello, ``{Connecting early and late epochs by f(z)CDM
  cosmography},'' \href{http://dx.doi.org/10.1088/1475-7516/2019/12/008}{{\em
  JCAP} {\bfseries 12} (2019) 008},
  \href{http://arxiv.org/abs/1910.09975}{{\ttfamily arXiv:1910.09975
  [astro-ph.CO]}}.

\bibitem{Schutz:1986gp}
B.~F. Schutz, ``{Determining the Hubble Constant from Gravitational Wave
  Observations},'' \href{http://dx.doi.org/10.1038/323310a0}{{\em Nature}
  {\bfseries 323} (1986) 310--311}.

\bibitem{Holz:2005df}
D.~E. Holz and S.~A. Hughes, ``{Using gravitational-wave standard sirens},''
  \href{http://dx.doi.org/10.1086/431341}{{\em Astrophys. J.} {\bfseries 629}
  (2005) 15--22}, \href{http://arxiv.org/abs/astro-ph/0504616}{{\ttfamily
  arXiv:astro-ph/0504616}}.

\bibitem{Chen:2017rfc}
H.-Y. Chen, M.~Fishbach, and D.~E. Holz, ``{A two per cent Hubble constant
  measurement from standard sirens within five years},''
  \href{http://dx.doi.org/10.1038/s41586-018-0606-0}{{\em Nature} {\bfseries
  562} no.~7728, (2018) 545--547},
\href{http://arxiv.org/abs/1712.06531}{{\ttfamily arXiv:1712.06531
  [astro-ph.CO]}}.

\bibitem{DiValentino:2018jbh}
E.~Di~Valentino, D.~E. Holz, A.~Melchiorri, and F.~Renzi, ``{The cosmological
  impact of future constraints on $H_0$ from gravitational-wave standard
  sirens},'' \href{http://dx.doi.org/10.1103/PhysRevD.98.083523}{{\em Phys.
  Rev.} {\bfseries D98} no.~8, (2018) 083523},
\href{http://arxiv.org/abs/1806.07463}{{\ttfamily arXiv:1806.07463
  [astro-ph.CO]}}.

\bibitem{Palmese:2019ehe}
A.~Palmese {\em et~al.}, ``{Gravitational Wave Cosmology and Astrophysics with
  Large Spectroscopic Galaxy Surveys},''
\href{http://arxiv.org/abs/1903.04730}{{\ttfamily arXiv:1903.04730
  [astro-ph.CO]}}.

\bibitem{Abbott:2017xzu}
{\bfseries LIGO Scientific, Virgo, 1M2H, Dark Energy Camera GW-E, DES, DLT40,
  Las Cumbres Observatory, VINROUGE, MASTER} Collaboration, B.~P. Abbott {\em
  et~al.}, ``{A gravitational-wave standard siren measurement of the Hubble
  constant},'' \href{http://dx.doi.org/10.1038/nature24471}{{\em Nature}
  {\bfseries 551} no.~7678, (2017) 85--88},
\href{http://arxiv.org/abs/1710.05835}{{\ttfamily arXiv:1710.05835
  [astro-ph.CO]}}.

\bibitem{DiValentino:2017clw}
E.~Di~Valentino and A.~Melchiorri, ``{First cosmological constraints combining
  Planck with the recent gravitational-wave standard siren measurement of the
  Hubble constant},'' \href{http://dx.doi.org/10.1103/PhysRevD.97.041301}{{\em
  Phys. Rev.} {\bfseries D97} no.~4, (2018) 041301},
\href{http://arxiv.org/abs/1710.06370}{{\ttfamily arXiv:1710.06370
  [astro-ph.CO]}}.

\bibitem{Nissanke:2013fka}
S.~Nissanke, D.~E. Holz, N.~Dalal, S.~A. Hughes, J.~L. Sievers, and C.~M.
  Hirata, ``{Determining the Hubble constant from gravitational wave
  observations of merging compact binaries},''
  \href{http://arxiv.org/abs/1307.2638}{{\ttfamily arXiv:1307.2638
  [astro-ph.CO]}}.

\bibitem{Palmese:2020aof}
{\bfseries DES} Collaboration, A.~Palmese {\em et~al.}, ``{A statistical
  standard siren measurement of the Hubble constant from the LIGO/Virgo
  gravitational wave compact object merger GW190814 and Dark Energy Survey
  galaxies},'' \href{http://arxiv.org/abs/2006.14961}{{\ttfamily
  arXiv:2006.14961 [astro-ph.CO]}}.

\bibitem{Yu:2020vyy}
J.~Yu, Y.~Wang, W.~Zhao, and Y.~Lu, ``{Hunting for the host galaxy groups of
  binary black holes and the application in constraining Hubble constant},''
  \href{http://arxiv.org/abs/2003.06586}{{\ttfamily arXiv:2003.06586
  [astro-ph.CO]}}.

\bibitem{Borhanian:2020vyr}
S.~Borhanian, A.~Dhani, A.~Gupta, K.~Arun, and B.~Sathyaprakash, ``{Dark Sirens
  to Resolve the Hubble-Lema\^itre Tension},''
  \href{http://arxiv.org/abs/2007.02883}{{\ttfamily arXiv:2007.02883
  [astro-ph.CO]}}.

\end{thebibliography}\endgroup

\end{document}